\documentclass{JASSS2arxiv}
	
\title{Validating argument-based opinion dynamics with survey experiments}

\reviewcopy{true} 

\author[1]{Sven Banisch}
\affil[1]{Karlsruhe Institute of Technology, Institute for Technology Futures}

\author[2]{Hawal Shamon}
\affil[2]{Forschungszentrum Jülich}

\email{Corresponding author email here}

\usepackage{natbib}
	\setcitestyle{authoryear,round,aysep={}}

\newcommand{\rem}[1]{}
\usepackage{nameref}
\usepackage{float}

\usepackage{hyperref}

\definecolor{darkgreen}{rgb}{0.0, 0.4, 0}
\newcommand{\hawal}[1]{{\color{red}{#1}}}

\definecolor{darkblue}{rgb}{0.0, 0.7, 0}

\definecolor{darkred}{rgb}{0.7, 0.0, 0}
\newcommand{\revision}[1]{{\color{black}{#1}}}

\begin{document}
\maketitle 



\begin{abstract}

The empirical validation of models remains one of the most important challenges in opinion dynamics. In this contribution, we report on recent developments on combining data from survey experiments with computational models of opinion formation. We extend previous work on the empirical \revision{assessment} of an argument-based model for opinion dynamics in which biased processing is the principle mechanism. 
While previous work \citep{Banisch2021biased} has focused on calibrating the micro mechanism with experimental data on argument-induced opinion change, this paper concentrates on \revision{the macro level} using the empirical data gathered in the survey experiment. For this purpose, the argument model is extended by an external source of balanced information which allows to control for the impact of peer influence processes relative to other noisy processes. We show that surveyed opinion distributions are matched with a high level of accuracy in a specific region in the parameter space, indicating an equal impact of social influence and external noise. More importantly, the estimated strength of biased processing given the macro data is compatible with those values that achieve high likelihood at the micro level. The main contribution of the paper is hence to show that the extended argument-based model provides a solid bridge from the micro processes of argument-induced attitude change to macro level opinion distributions. 
Beyond that, we review the development of argument-based models and present a new method for the automated classification of model outcomes.
\end{abstract}

\begin{keywords}
opinion dynamics, \revision{validation, empirical confirmation, survey experiments,} parameter estimation, argument communication theory, computational social science
\end{keywords}

\parano{}



\section{Introduction}

Opinion dynamics is a field that develops theoretical models of collective opinion processes to understand the mechanisms behind the emergence of consensus, polarization and conflict.
It uses agent-based computational models (ABMs) to simulate the evolution of opinions in a population of artificial agents.
These agents are placed in a social environment typically consisting of an interaction network defining who can interact with whom.
In the course of a simulation neighboring agents interact and exchange opinions according to some simple rules.
Opinion dynamics studies the properties of these complex dynamical systems to identify basic mechanisms behind different collective phenomena from consensus to different forms of polarization.

A lot of modeling work in the last two decades has been inspired by the so-called "puzzle of polarization" frequently referring to \cite{Abelson1964mathematical} and \cite{Axelrod1997dissemination}.\footnote{\cite{Abelson1964mathematical}: "Since universal ultimate agreement is an ubiquitous outcome of a very broad class of mathematical models, we are naturally led to inquire what on earth one must assume in order to generate the bimodal outcome of community cleavage studies." \cite{Axelrod1997dissemination}: "If people tend to become alike in their beliefs, attitudes, and behavior when they interact, why do not all such differences eventually disappear?"}
The motivating question has been: How does a population with moderate initial opinions diverge into groups of agents that strongly support opposing views?
Early models \citep{French1956formal,DeGroot1974reaching, Friedkin1990social} implementing positive social influence by which opinions assimilate in interaction predict consensus whenever the interaction network is a single connected component.
Research in the last 20 years has revealed quite a few mechanisms that may solve the puzzle of persistent opinion plurality including bounded confidence \citep{Deffuant2000mixing,Hegselmann2002opinion} introduced in the two papers to which this special issue is in some sense devoted.
Other models targeting bi-polarization dynamics draw on more sophisticated forms of homophily \citep{Carley1991theory,Maes2013differentiation}, negative influence \citep{Jager2005uniformity,Baldassarri2007dynamics,Flache2011small,Maes2014cultural}, opinion reinforcement \citep{Martins2008continuous,Banisch2019opinion}, biased assimilation \citep{Dandekar2013biased,Mueller2018cognitive,Banisch2021biased}, and combinations of those.
Most of these models are covered by the review of \cite{Flache2017models} in this Journal and the social influence wiki \citep{social-influence-wiki} initiated by its authors.

Nowadays, 20 years later, many models exist that provide \emph{possible explanations} of collective bi-polarization and we are facing the problem to select the most relevant mechanisms given more specific questions.
The field is ripe to take a further step beyond the mere theoretical exploration of how qualitatively different idealized macro phenomena, such as consensus and polarization, may arise from different basic micro-level assumptions.
As a matter of fact, there is a great need for more realistic models. 
Especially in an era where collective communication is more and more engineered, where social network algorithms guide what becomes visible to whom, simulation tools are needed to rigorously inquire and predict the potential impact of algorithmic filters and platform choices.
Recent work on algorithmic personalization and filter bubbles has shown that different combinations of micro mechanisms may lead to conflicting predictions concerning the impact of algorithm-induced homophily on opinion dynamics \citep{Maes2015will,Keijzer2022complex}.
This is a big obstacle when using model results as a basis for science-based policy recommendations.
In order to draw rigorous conclusions from simulation models we have to decide which of these principal mechanisms are most prevalent given an empirical case.
 
In order to advance towards \emph{applied opinion dynamics}, the empirical validation of opinion models remains a major challenge for the field \citep{Sobkowicz2009modelling,Flache2017models}.\footnote{See \cite{Chattoe2022if} and \cite{Keijzer2022if}\revision{ as well as \cite{ChattoeBrown2022today}, \cite{carpentras2023failing} and \cite{Neumann2023challenge} for recent discussions} on the topic.}
Data usually does not fit well with the idealized world of opinion models and in fact there are only few topics on which real opinions compare to the stylized pattern of bi-polarization that emerges in most of the models \citep{Duggins2017psycologically}.
In this paper, we aim to advance the state of the art of empirical validation in opinion dynamics by combining a survey experiment \citep{Shamon2019changing} on argument persuasion with argument-based models of opinion formation \citep{Maes2013differentiation,Banisch2021argument,Taillandier2021introducing}.
The experiment provides micro-level data on attitude change and macro-level data on opinions with respect to six different technologies for electricity production.
This opens the possibility to calibrate the micro-level mechanisms of the ABM and to compare resulting opinion distributions to real opinions on the same topics.
The main objective is to empirically interrogate argument communication theory (ACT) \citep{Maes2013differentiation} so that models developed within this paradigm can be confidently applied to real problems such as the impact of online policies on opinion dynamics.

To achieve that, the present paper extends previous work that introduced biased argument processing into argument communication models and showed that this mechanism can explain experimentally observed opinion changes better than previous models \citep{Banisch2021biased}. 
While this first paper has focused on calibrating the micro mechanism with experimental data on argument-induced opinion change, \revision{the present} paper concentrates on the macro-level comparison of model outcomes to the survey data gathered in the experiment.
The main contribution of this paper, is to show that the argument communication model with biased processing also reproduces macro-level data on opinions with high accuracy if we control for the impact of social influence. Namely, we extend the model by assuming a certain level of unbiased external information that supplies the system with random arguments (as opposed to arguments brought up by peers). We analyze the impact of this form of noise on the model behavior and show that in a regime of moderate biased processing and a relatively high level of noise opinion data is matched remarkably well by its stationary distribution.

The model hence provides a consistent link between empirical observations at the micro and the macro level.
It provides an empirically-grounded explanation that bridges from individual patterns of attitude change to the resulting distributions of surveyed opinions.
We believe that this is an important step towards empirically validated argument-based models which prepares them for more specific applications.


The paper is structured as follows. The next section presents background and the current state of research. We discuss binary and continuous opinion models as the two major traditional model classes, and sketch the development of argument-based opinion dynamics as a combination of those. \revision{The next section 
introduces the model and provides details on the computational analysis. Section 4 provides results regarding the general behavior of the model in different regions of the parameter space. Section 5 presents the overall validation approach, discusses associated terminology and describes the survey experiment. Section 6 
finally revisits previous results on model calibration, provides the results regarding the comparison of model outcomes to the empirical opinion distributions, and compares the two.} We conclude with a discussion on validation in opinion dynamics and the potential contributions of this work.

    



\section{Argument communication theory: predecessors and theoretical development}

Opinion dynamics is an interdisciplinary endeavor that has attracted researchers from physics, computer science and mathematics as well as sociologists, political science and communication scholars. At a very basic level, one can distinguish different models with respect to what they treat as an agent’s opinion. While the physics community has mostly concentrated on binary (or discrete) state models in which the opinion is a single nominal variable, the idea that an opinion is a metric variable on a continuous opinion scale from -1 to 1 is prevailing in the social simulation community. Argument communication theory (henceforth ACT) combines aspects of both binary choice and continuous space models and therefore we provide a brief and selective overview of modeling work within these two paradigms. A more encompassing review on both model classes has been provided in \cite{Sirbu2017opinion}. 
See also \cite{Lorenz2007continuous,Flache2017models} for reviews on continuous opinion models and \cite{Galam2008sociophysics,Castellano2009statistical} for physics-inspired models. 


\subsection{Binary and continuous opinion dynamics}

\subsubsection{Binary state opinion models}

In binary state models agents are characterized by a single binary variable, say $o_i \in \{0,1\}$.
In the interaction process, an agent $i$ is chosen and updates its opinion according to opinions in its neighborhood. Models mainly differ with respect to how this update is conceived. 
The most simple and well-studied binary opinion model -- the voter model -- originated in theoretical biology as a model for the spatial conflict of two species \citep{Kimura1964stepping}. 
In the voter model, a single neighbor $j$ is chosen at random out of the neighbors set of $i$ and $i$ copies the state of $j$ (i.e. $o_i \leftarrow o_j$) \citep{Holley1975ergodic,Banisch2016springer}. In majority rule models \citep{Galam1986majority,Chen2005majority,Galam2008sociophysics}, $i$ sees the states of all neighbors and updates its opinion by following the local majority. These models are therefore closely related to early threshold models \citep{Schelling1973hockey,Granovetter1978threshold} where opinion update takes place when a certain fraction of neighbors assumes an alternative state. More complex forms of frequency dependence have been studied with so-called non-linear voter models \citep{Schweitzer2009nonlinear} or the Sznajd model \citep{Sznajd2000opinion}. Notice finally that also the social impact model introduced in 1990 by \cite{Nowak1990private} falls into the category of binary models.

The main reason for which binary state opinion dynamics has attracted so much attention from the physics community is their analogy to spin systems. Networks of agents that switch between two opinion states by following the choices of neighbors resemble physical systems of ferromagnetically coupled Ising spins \citep{Ising1925beitrag}. For this reason, the concepts and tools of statistical physics can be applied to study -- often analytically -- the dynamical behavior of a model \citep{Lewenstein1992statistical,Castellano2009statistical}. One very central concept is the so-called Hamiltonian that assigns an energy to each possible configuration of the system according to the network of social coupling. A computational model that implements local opinion alignment can then be seen as a relaxation dynamics approaching the (local or global) minima in this energy landscape. A second important idea that the engagement of physicists with opinion dynamics models brought into the field is that of a phase transition, and associated critical points. A phase transition indicates that the behavior of a model undergoes a qualitative change as model parameters change. In binary opinion dynamics this often means a transition from consensus (all agents aligned) to a disordered state under increasing levels of noise \citep{Holyst2000phase, Nowak2020symmetrical}, contrarian agents \citep{Galam2004contrarian,Banisch2014microscopic,Krueger2017conformity} or zealots \citep{Mobilia2003does,Crokidakis2015inflexibility}. Especially close to the transition point the models often exhibit very interesting long-lasting mesoscopic patterns such as non-stationary local clusters of agents with aligned opinions \citep{Schweitzer2009nonlinear}. It is noteworthy, that the relation between opinion dynamics and statistical mechanics has been productive in both directions. Problems of social dynamics have motivated significant research on how to tackle heterogeneous and complex networks with physics tools such as mean field approximations \citep{Sood2005voter,Vazquez2008analytical} and has played a big role in the development of pair and higher-order approximation techniques \citep{Schweitzer2009nonlinear,Gleeson2013binary}.

Binary state dynamics over complex networks can be seen as a blueprint of a complex dynamical system and therefore these models have seen applications in all fields that have embraced the turn to complexity in the last decades. The wide applicability derives from the fact that the two possible states are open to many interpretations, including the absence or presence of biological species as in the original voter model by \cite{Kimura1964stepping}. In the context of opinion dynamics, they can relate to beliefs and opinions, but also to alternative behaviors as in the literature on complex contagion \citep{Centola2007complex,Ugander2012structural} and games on networks \citep{Szabo2007evolutionary,Galeotti2010network}.


\subsubsection{Continuous opinion dynamics}

While the previous class of models originated in theoretical biology, the origins of continuous opinion dynamics can be traced back to early research in mathematical social psychology on consensus formation in small groups \citep[see the first two chapters in][for a historical perspective]{Friedkin2011social}. 
Here an agent's opinion is represented as a continuous variable which typically represents a degree of favor versus disfavor (i.e. an attitude $o_i = [-1,1]$), or a subjective probability or belief ($o_i \in [0,1]$).
In these early models a crucial construct has been the social influence matrix $W$ that encodes the relative influence an individual $j$ exerts on any other individual $i$.
In the dynamical process, the new opinion of an agent is given as the weighted average of an agent's own current opinion (weighted by $w_{ii}$) and those of its neighbors ($o_i \leftarrow \sum_j w_{ij} o_j$). This is referred to as positive or assimilative social influence.
Formally, this repeated averaging process can be written as a linear system $o^{t+1} = W o^t = W^t o^1$ where $o$ is the evolving $N$-dimensional vector with the opinions of all agents \citep{French1956formal}.
Such systems converge to a consensual final state (i.e. $o_i = o_j \ \forall i,j$) whenever the matrix of interpersonal influences $W$ consists of a single connected component \citep{DeGroot1974reaching}.

Bounded confidence models \citep{Deffuant2000mixing, Hegselmann2002opinion} have been invented against this background to show how multiple opinion groups can persist under social influence dynamics. The idea is simple: two agents with opinions $o_i$ and $o_j$ influence one another only when they are already close enough in opinion space, that is, when the distance between $o_i$ and $o_j$ is below a certain confidence threshold. \cite{Hegselmann2002opinion} describe very well that this extension to social influence network models formally leads to a non-linear system in which the influence matrix $W$ changes through time. Since then, most work within the continuous opinion paradigm is based on computer simulation \citep[see][for notable exceptions]{Friedkin2015problem, Friedkin2016network}.
If the threshold is low enough, the influence network features isolated groups of individuals within a certain range of opinions that converge to a group consensus independently from other groups. The number of groups depends on the confidence threshold in non-trivial ways\footnote{This point has been emphasized in a recent talk by R. Hegselmann at the annual meeting of the German Physical Society in Regensburg 2022.},
but the model can lead to complete fragmentation into many opinion groups, to the persistence of two opposing opinion groups, or to consensus. 

Like in binary opinion dynamics many different social and psychological assumptions have been integrated into the models in subsequent work. First, more realistic forms of relative homophily take into account that the probability of interaction depends on how many similar agents are available \citep{Carley1991theory, Maes2013differentiation, Baumann2020modeling}. Second, negative social influence has been proposed as an additional mechanism by which agents differentiate from other agents that are already different when they interact \citep{Jager2005uniformity,Baldassarri2007dynamics,Flache2011small}. The repulsive forces implemented by negative influence may somewhat trivially lead to extreme bi-polarization, but the empirical relevance of the mechanism is disputed \citep{Takacs2016discrepancy}.
Another mechanism that leads to extreme bi-polarization is opinion reinforcement by which pairs of agents strengthen their conviction if they are on the same side of the attitude scale \citep{Martins2008continuous,Banisch2019opinion}.
There are different processes that may lead to such an opinion reinforcement including argument communication under homophily \citep{Maes2013differentiation,Maes2013short} \citep[cf.][par. 2.67 ]{Flache2017models}, social feedback \citep{Banisch2019opinion,Gaisbauer2020dynamics} and contagion \citep{Lorenz2021individual}. In terms of micro-level justification, opinion reinforcement can therefore draw on a rich body of psychological research on group polarization \citep{Myers1976group,Sunstein2002law} as well as on neuroscientific experiments on social reward processing \citep[cf.][]{Banisch2022modelling}. However, a recent experiment aimed at a direct measurement of opinion reinforcement through social approval has been inconclusive \citep{Sarkozi2022effects}.
Finally, also biased processing -- the central mechanism in this paper -- has been introduced in continuous state models \citep{deffuant2007propagation,Dandekar2013biased,Lorenz2021individual}. The existence of cognitive biases in the processing of information has been proven to be a robust mechanism across various empirical experiments on different issues \citep[e.g.,][]{Taber2006motivated,Taber2009motivated,Druckman2011framing,Corner2012uncertainty,Teel2006evidence} including the one used in this paper \citep{Shamon2019changing}.

It is noteworthy that models usually implement combinations of these core mechanisms and study how the model outcomes are affected by varying the mixture.
The biased assimilation model by \cite{Dandekar2013biased}, for instance, combines biased processing with homophily to generate bi-polarization. Other researchers have started to systematically address the micro-macro problems involved when drawing societal level conclusions from competing micro assumptions \citep{Maes2015will, Keijzer2022complex}. This research has shown, for instance, that filter bubbles and increasing personalization (modeled as homophily) lead to completely different conclusions depending on whether they are combined with argument-based opinion exchange or negative influence. Several authors have started to increase the psychological realism of models by more explicitly drawing on established psychological theories within a continuous opinion setting \citep{Duggins2017psycologically, Banisch2019opinion, Lorenz2021individual}. 
The model by \cite{Duggins2017psycologically}, for instance, integrates positive and negative social influence, conformity, distinction and commitment to previous beliefs as well as social networks to show that increased micro-level complexity is needed to generate realistic opinion distributions characterized by strong diversity. \cite{Lorenz2021individual} advance towards model synthesis -- the second main challenge identified in the review by \cite{Flache2017models} -- by drawing on a generalized attitude change function derived as an attempt to synthesize different psychological theories of attitude change \citep{Hunter2014mathematical}.


\subsection{Argument communication theory (ACT)}

ACT has been introduced in \cite{Maes2013short} and \cite{Maes2013differentiation} as a possible explanation for the emergence of opinion bi-polarization that does not draw on negative influence. The models combine aspects from binary opinion dynamics and continuous models by relying on a two-layered concept of opinion. That is, an agents' opinions ($o_i$) is assumed to be determined by an underlying string of binary arguments ($\vec{a}_i$) that may support a positive or a negative evaluation of an attitude object. Processes of information exchange in social interaction take place at the lower level of arguments by adopting arguments from peers. Opinions follow from that and change whenever a new argument is obtained. But opinions also become functional in the repeated exchange process as they structure lower-level information uptake by guiding partner selection (homophily) \citep{Maes2013differentiation} or opinion revision (biased processing) \citep{Banisch2021biased}.

\subsubsection{Original model by \cite{Maes2013differentiation}}

ACT has been inspired by psychological work on group polarization in the 1970ies and 80ies \citep{Myers1976group,Vinokur1978depolarization,Isenberg1986group} that observed that discussions may reinforce initial opinions of a group \citep[see also][]{Sunstein2002law}.
At that time, negative influence had become a frequent modeling choice in order to model a process of increasing divergence and the emergence of two increasingly opposing opinion camps at the extremes of an opinion scale. The model by \cite{Maes2013differentiation} showed that repeated processes of argument exchange in which agents locally assimilate may lead to polarization dynamics under homophily. This is possible through a more complex multi-layered conception of opinions as attitudes that rely on an underlying set of pro and con arguments. The argument exchange mechanism acts on the underlying level of arguments. But homophily acts at the upper layer of opinions defined as the number of pro versus con arguments. Under homophily, opinions act as social filters so that agents that already hold many pro (con) arguments will encounter other agents holding other pro arguments that further support a positive (negative) stance.

In their model, the number of arguments is relatively large, 30 pro and 30 con arguments. But agents can only "remember" a subset of 10 salient arguments which is actualized in interaction. Arguments are ranked according to their recency. If an agent receives a new argument from an interaction partner, that argument is activated ($a_{ik}=1$) and ranked first in recency. In turn, another argument of least recency is dropped ($a_{ik} = 0$). In this way, the model accounts for limited memory capacities and a higher accessibility of recent information. Opinions are then defined by the number of pro and con arguments that are currently salient. 

In \cite{Maes2013differentiation} and subsequent papers \citep{Maes2013short,Maes2015will,Keijzer2022complex} homophily is implemented as biased partner selection following earlier work by \cite{Carley1991theory}. It is a relative conception of homophily. First, an agent $i$ is chosen and then the interaction partner $j$ is drawn from all other agents with a probability that depends in a non-linear way on the opinion similarity between $i$ and $j$. The degree of favoring the most similar others is governed by a free parameter $h$ and the system polarizes if $h$ becomes large. As opposed to bounded confidence where the interaction probability of two agents $i$ and $j$ depends only on the opinion of the two involved agents, in these works the interaction probabilities depend on the opinions of all agents in the population. While this is plausible compared to the hard threshold of bounded confidence, it implicitly assumes that the opinions of all agents are known at each step. Moreover, the population-relative interaction probabilities have to be recomputed at each step which is very costly from the computational point of view.

\subsubsection{Model simplifications}

Against this background, it has been shown in \cite{Banisch2021argument} that the qualitative behavior of the Mäs-Flache model is preserved under more simple choices regarding the number of arguments, the exchange mechanism and homophily. In their model only 3 pro and 3 con arguments are used. Agents can either believe that an argument is true ($a_{ik} = 1$) or false ($a_{ik} = 0$). Opinions are then defined as in \cite{Maes2013differentiation} as the number of pro versus con arguments that an agent beliefs to be true. In the interaction process, two agents, a sender and a receiver, are chosen at random and the receiver copies a randomly chosen argument from the sender. Without homophily this corresponds to a multi-dimensional voter model (see above) where agents converge to a common state independently along each dimension. While all agents converge to a common argument string (and hence opinion) without homophily, the system will polarize under strong opinion homophily. This basic polarization dynamics is preserved if the relative homophily of the original model is replaced with bounded confidence.

The introduction of ACT of bi-polarization in \cite{Maes2013differentiation,Maes2013short} has also inspired new socio-physics models of opinion dynamics such as the M model by \cite{LaRocca2014influence}. In this model, the layer of arguments is not explicitly modeled. 
In our reading of the theory it is not an ACT model. 
But the two mechanism of persuasion and compromise that the model uses to define opinion change mimic the opinion changes that would be observed under argument exchange. Enabling the application of statistical physics, this approach is very useful for a better and more rigorous understanding of the statistical properties and phase transitions of ACT models that remain with a more complex structure of underlying arguments. 


\subsubsection{Interacting arguments}

One great benefit of increasing complexity at the level of individual opinion is the possibility to more explicitly represent issues of real debate as well as involved argumentation processes. In \cite{Taillandier2021introducing} a model has been presented in which opinions on vegetarian diet are represented by an underlying argument network. The model formalizes \cite{Dung1995acceptability}'s argumentation graphs in which arguments may attack other arguments. In terms of interaction and homophily the paper follows \cite{Maes2013differentiation}. But using the theoretical concepts of argumentation graphs it implements argument choice by the sender and acceptance by the receiver based on consistency computations on the argument attack network. This can be seen as a form of biased processing. The model studied in \cite{Taillandier2021introducing} is empirically-informed by real arguments and attack relations, drawing on a set of 145 arguments on vegetarian diet. 
Closely following the computational design of \cite{Maes2013differentiation}, the study shows that interacting arguments have an impact on the consensus-polarization transition caused by increasing homophily.

\subsubsection{Multiple interrelated issues}


Another attempt to more realistically capture the complexity of real debates has been made in \cite{Banisch2021argument}. There are no links between arguments, but a bipartite network that links arguments to multiple issues of opinion. These cognitive-affective networks entail evaluative associations in a way closely related to psychological theories of attitude structure and associated measurements \citep{Fishbein1962ab,Fishbein1963investigation,Ajzen2001nature}. In this setting, arguments may be relevant to more than one issue such as trading one versus the other. Based on a simplified argument exchange process (see above), the model accounts for polarization in terms of ideological alignment or opinion sorting on various issues. This kind of opinion alignment is a robust empirical fact, for instance, political dimensions such as "left" versus "right" are only meaningful only because of specific patterns of attitudinal correlations \citep{Laver1992measuring,Laver2014measuring,Olbrich2021rise}.
Also for this model attempts to derive realistic argument-opinion relations from textual data have been made \citep{Willaert2022tracking}.

\subsubsection{Incorporation of biased processing}

In empirical research, there are various randomized experiments that have investigated the influence of the exchange of arguments on opinions and whose design is very close to the conceptualization of the ACT  \citep[e.g.,][]{Taber2006motivated,Taber2009motivated,Druckman2011framing,Corner2012uncertainty,Teel2006evidence,Shamon2019changing}.
These empirical experiments have in common that participants are first asked about their opinion towards a certain issue under investigation. Then, they are sequentially exposed to arguments in favor of or against the issue and, finally, asked one more time on their opinion on the investigated issue. In this way, the empirical experiments measure opinion changes among participants as a result of exposure to pro- and con-arguments. In contrast to the ACT, however, the opinion change is not measured again after each argument exposure, but at the end of the confrontation with all arguments. Even more important, these experiments find empirical evidence for a cognitive mechanism, called biased processing, that is not addressed by ACT-assumption. Biased processing refers to a person's tendency to inflate the quality of arguments that are compatible with his or her existing opinion on an issue whereas the quality of those arguments that speak against a person’s prevailing opinion are downgraded. This empirical robust cognitive mechanism challenges ACT's assumption of argument adaption at a constant rate independent of the current opinion. 

Biased processing has been incorporated into ACT in \cite{Banisch2021biased}. The model operates with 4 pro and 4 con arguments which are copied in interaction. However, the probability to accept a new argument depends in a non-linear way on the current opinion of the receiver. This is modeled by a soft-max (or Fermi) function that contains a parameter $\beta$ which accounts for the strength of biased processing. If $\beta$ is zero, arguments are accepted with equal probability independent of the opinion. If $\beta$ is large arguments that speak against the current opinion are rejected whereas arguments that further support the current stance are accepted with a probability close to one. 
In \cite{Banisch2021biased} $\beta$ has been estimated from experimental data (see below) and moderate values ($\beta \approx 0.5$) have been found. The computational analysis of the model has shown that biased processing has a strong effect on the behavior of the argument model. First, as soon as $\beta > 0$, the stability of moderate consensus is lost and group converge to one or the other extreme on the opinion scale. Second, strong biased processing ($\beta > 1$) may lead to a meta-stable state of bi-polarization even in the absence of homophily. 
In this paper, we study this version of the argument model with noise and provide a more detailed model description in the \revision{next} section.


\section{Argument model with biased processing and noise}
\label{sec:model}

In this paper, we extend the model of \cite{Banisch2021biased} by introducing an external source of information that supplies the system with random arguments. 
\revision{In this section, we describe the model, show a series of paradigmatic model realizations, and describe what we treat as a model outcome for comparison to data.
We also provide details on the computational strategy and the implementation.
}


\subsection{Model description}

\revision{We model a system of $N$ agents that exchange arguments about a single opinion item in repeated interaction. If not stated otherwise we will use $N = 500$ agents. Here we describe the iterative process following the order in which the different steps are performed. We start with the opinion structure.}

\begin{figure}[ht]
	\centering
	\includegraphics[width=0.4\linewidth]{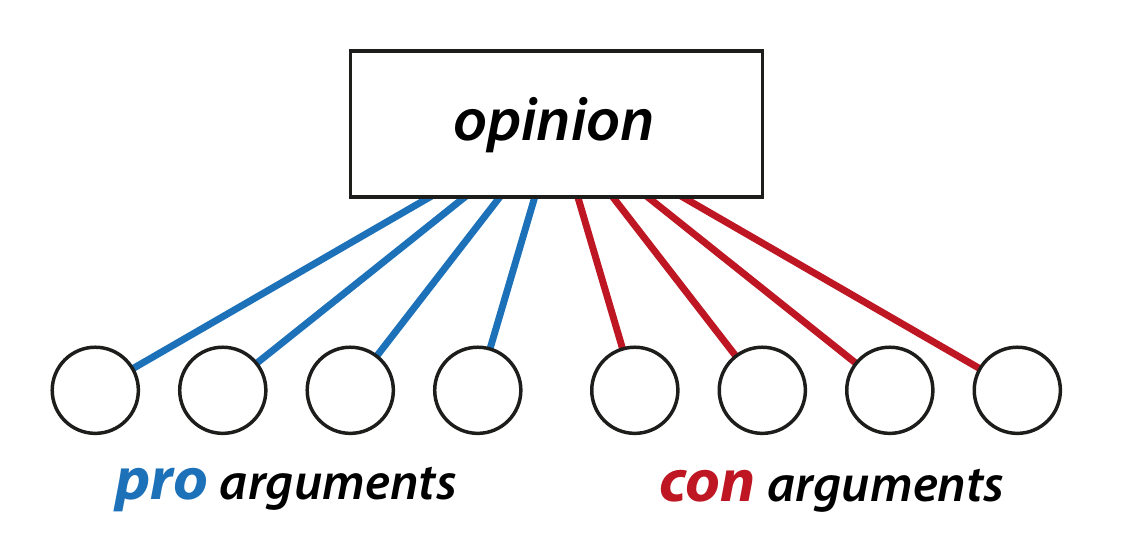}
	\caption{Two-layered opinion structure. Opinions are defined by an underlying set of pro and con arguments.}
	\label{fig:opinionstructure}
\end{figure}

\revision{
\textbf{Opinion structure.} Models within the framework of ACT rely on a two-layered conception of opinions. They assume that the opinion $o_i$ of an agent $i$ is determined by an underlying layer of pro and con arguments that the agents holds. In our model,} each agent is endowed with a string of $K$ binary variables which we call arguments or beliefs. We denote the argument string of a single agent $i$ as ($\vec{a}_{i} = (a_{i1}, \ldots, a_{iK}) \in \{0,1\}^K$). Homogeneously for the entire population, we assume that the first $K/2$ arguments are pro arguments and the latter \revision{half ($k > K/2$)} are counter arguments. See Figure \ref{fig:opinionstructure}. \revision{For further convenience we introduce a $K$-dimensional vector of evaluations $e_k$ where the first $K/2$ elements are $+1$ (pro arguments) and the second $K/2$ arguments are $-1$ (con arguments) \citep[see][for a psychological motivation]{Banisch2021argument}.
An agent's opinion, defined as the number of pro versus con arguments ($n_+$ and $n_-$), can then be defined as
\begin{equation}
	o = \sum_{k=1}^{K} a_k e_k = n_{+} - n_{-}
	\label{eq:opinion}
\end{equation}
}
Hence, if an agent believes a pro argument to be true ($a_{ik} = 1, k \leq K/2$), this will contribute an amount of $+1$ to a positive opinion ($o_i$). Vice versa, an argument $a_{ik} = 1, k > K/2 $ contributes an amount of $-1$ to a negative stance.
\revision{In this paper, we set the number of arguments to $K = 8$ to align the opinions in the computational model with the 9-point answer scale that was used in the survey experiment. This avoids distortions that may arise when scales of different ranges are being applied in both the empirical measurement and the ABM \citep[cf. e.g.][]{carpentras2023psychometric}.}

\revision{\textbf{Initial conditions.}} At start ($t=0$), the system is initialized by assigning random arguments to all the agents. That is, for each single argument $a_{ik}$ there is a fifty-to-fifty chance that $a_{ik} = 0$ or $a_{ik} = 1$. Consequently, the initial distribution of opinion follows a binomial distribution as shown on the left-hand side in Figure \ref{fig:run1}. 
\revision{Throughout the paper, we focus on the stationary dynamics of the model reducing the impact of different initial conditions.}

\revision{
\textbf{Partner selection and update schedule.}
Following \cite{Banisch2021biased} we use the following schedule to update the system. At each time step}, we draw $N/2$ random agent pairs (without replacement). \revision{All agent pairs have an equal probability to be chosen and their is no interaction network or homophily (random mixing).} The first agent $s$ is considered as a sender and the second agent as a receiver (denoted respectively as $r$). This means that in a single simulation step ($t \rightarrow t+1$) each agent is chosen either as a sender or a receiver. In other words, in our model implementation one time step corresponds to $N/2$ update events.

\revision{
\textbf{Argument exposure: social influence versus noise.}
In the original model by \cite{Banisch2021biased}, for each pair, the receiver $r$ is exposed to an argument randomly drawn from the argument string $\vec{a}_s$ of the sender.} That is, the articulated argument $arg_k = a_{sk}$ with $k$ drawn uniformly from $(1,\ldots,K)$. \revision{We refer to this as social influence condition. Here, we extend this model by assuming that $r$ receives an argument from an external source with a certain probability $\rho$.
This parameter allows to control for the impact of social influence ($1-\rho$) versus noise ($\rho$). Hence, f}or each pair, we first decide whether $r$ receives an argument from the sender $s$ or a random argument from an external source.  In the social influence condition (with probability $1-\rho$), we randomly choose an argument from $\vec{a}_s$. In the noise condition ($\rho$), we also randomly select a $k$ uniformly from $(1,\ldots,K)$, but assign a random binary value to $arg_k$. The receiver $r$ receives $arg_k$ and the decision to adopt the argument is equal in the two influence conditions, and subject to biased processing in both cases. 

\begin{figure}[ht]
	\centering
	\includegraphics[width=0.6\linewidth]{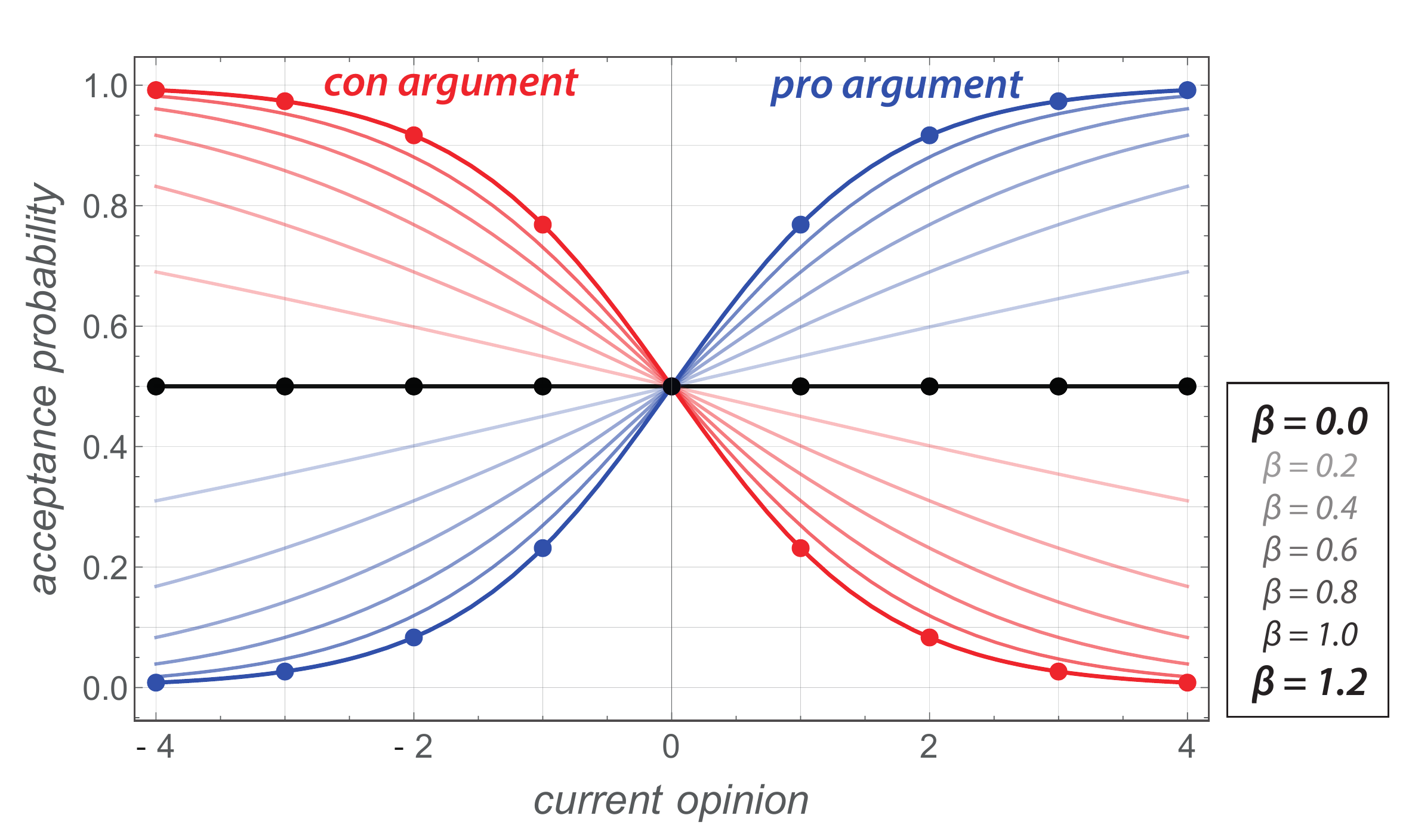}
	\caption{The receiver adopts an argument with the opinion-dependent probability $p_{\beta}(o_r)$. A counter argument (red) is favored if the current opinion is negative and is more likely rejected if $o_r > 0$. Vice versa for a pro argument (blue). \revision{Self-confirmatory} argument acceptance becomes more pronounced as $\beta$ increases (shade of respective color). Under unbiased adoption with $\beta = 0$ (black) arguments are accepted with probability $1/2$ independent of $o_r$.	}
	\label{fig:pAdopt}
\end{figure}

\revision{
\textbf{Argument adoption.}}
Under biased processing the probability that $r$ accepts the new argument $arg_k$ depends on $r$'s current opinion $o_r$ such that information that confirms the current opinion is accepted at a higher rate. The strength of this \revision{confirmation bias} -- sometimes referred to as "my-side" bias -- is governed by a second parameter $\beta$. 
We model this \revision{tendency to favor coherence over incoherence in the argument acceptance probabilities using a softmax or Fermi function. That is, an argument $arg_k$ is accepted with
\begin{equation}
   p_{\beta}(o_r,arg_k) = \frac{1}{1 + e^{-\beta o_r (2 arg_k -1)e_k}},
   \label{eq:biasedprocessing}
\end{equation}
where the term $(2 arg_k - 1)e_k$ encodes the evaluative direction of the argument $arg_k$ (pro versus con).}
To provide better intuition about \revision{this} function it is shown in Figure \ref{fig:pAdopt}. If $\beta = 0$ there is no \revision{confirmation} bias and all arguments are accepted with a chance of 50 percent ($p_{\beta} = 1/2$). There is no dependence on $r$'s current opinion. As $\beta$ increases a counter argument (red curves) has a high chance to be accepted by an agent with negative opinion whereas an agent with a positive opinion will more likely reject it. A pro argument (blue curves) is accepted with a probability higher than chance if the receiver has already a positive opinion ($o_r > 0$) whereas a receiver with a negative stance ($o_r < 0$) more likely rejects pro arguments. The Figure shows that this effect becomes relatively strong. For instance, consider $\beta = 1.2$ and $o_r = 1$. Such an agent will accept a pro argument with $p_{\beta} \approx 0.77$ but a counter argument only with probability $p_{\beta} \approx 0.23$.

\begin{figure}[ht]
	\centering
	\includegraphics[width=0.85\linewidth]{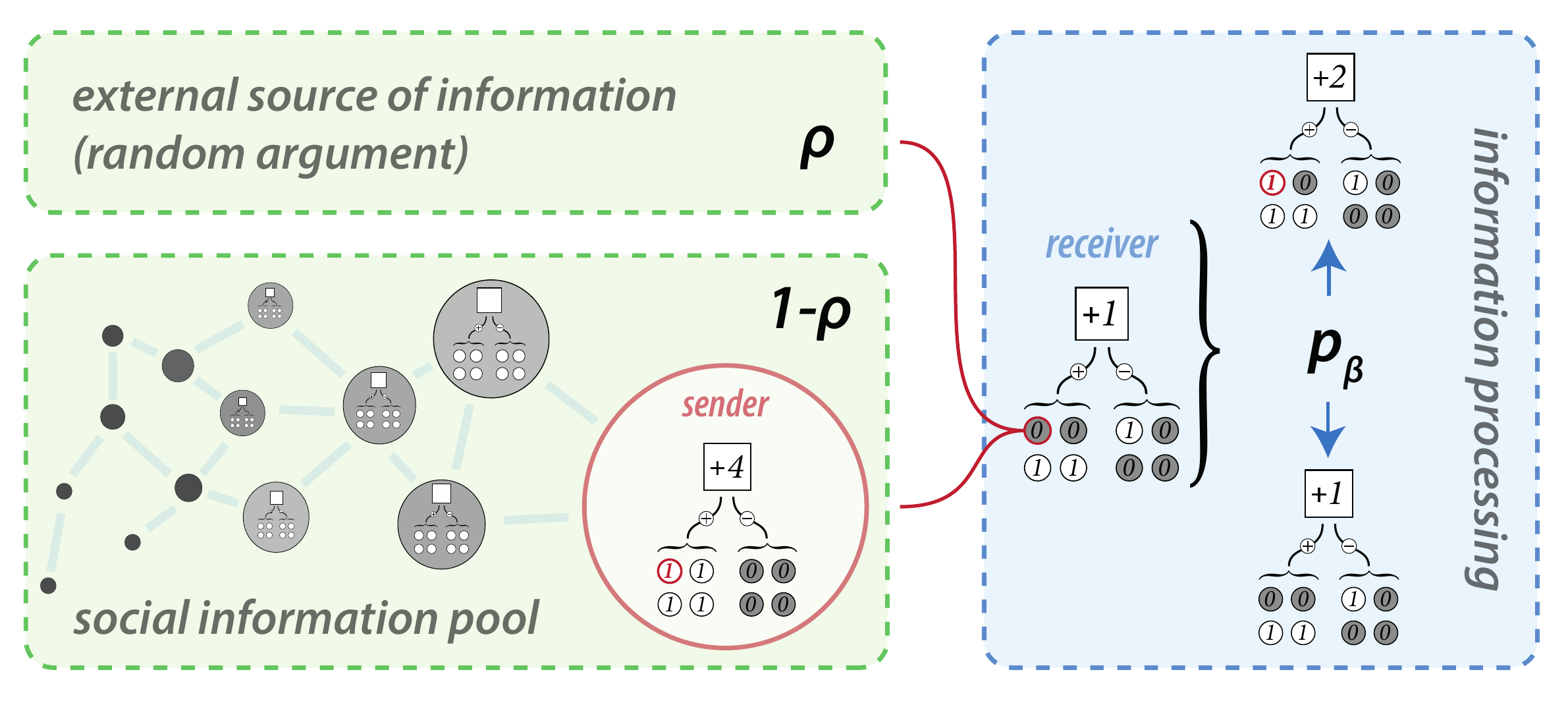}
	\caption{Illustration of the interaction process of the agent-based model. A random agent is chosen as the receiver. This agent receives an argument from another agent in the social pool (with probability $1-\rho$) or a random argument from an external source (with probability $\rho$). In both cases, the receiver evaluates the argument and adopts it with the bias-dependent probability $p_{\beta}$.
	}
	\label{fig:model}
\end{figure}

\revision{
\textbf{Model summary.}}
Figure \ref{fig:model} summarizes the argument communication model used in this paper. At each step, we randomly draw $N/2$ agent pairs and assign them the roles of sender and receiver $(s,r)$. For each pair we perform the following steps:
\begin{enumerate}
        \item[1.] random choice of an argument index $k$ in $(1,\ldots,K)$
        \item[2a.] social influence condition: with probability $1-\rho$ take $arg_k = a_{sk}$ from the sender $s$
        \item[2b.] external influence condition: with probability $\rho$ take $arg_k \in \{0,1\}$ with equal probability
        \item[3.] receiver $r$ accepts the argument $a_{rk} = arg_k$ with probability \revision{$p_{\beta}(o_r,arg_k)$}
        \item[4.] update of $r$'s opinion if the argument \revision{has changed}
\end{enumerate}

\revision{
\textbf{Simplifying assumptions.}}
Notice that in this paper we do not incorporate realistic social networks but rely on the complete graph as an underlying topology. We match agent pairs completely at random so that any pair is equally likely (random mixing). \revision{As opposed to e.g. \cite{Maes2013differentiation}, our model does not include biased partner selection \citep{Flache2017models} in form of homophily.}
There is also no particular mechanism behind choosing the argument $k$ (point 1), \revision{so to incorporate motivated reasoning.} The focus in this paper is on \revision{self-confirmatory information processing} and the impact of an unbiased external signal modeled as a form of noise that supplies random arguments to the system.
We differentiate a social influence from an external influence condition and the parameter $\rho$ decides on the respective probabilities. That is, $\rho$ determines the relative importance of peer influence versus external influence which we may consider as a very simple model for an unbiased media channel. \revision{In this regard, we do also not assume biases in information selection or attention which could be integrated by assuming that an agent more likely chooses confirmatory arguments \citep{Deffuant2023regular}.}
Notice finally that \revision{agents do not remember arguments held in the past} and that $r$'s opinion \revision{does not change (point 4) if the argument $arg_k$ confirms an argument $r$ already holds. In this case,} the argument string \revision{does not change} and $o_r$ is not affected.

\subsection{Paradigmatic simulation runs}

\begin{figure}[h]
	\centering
	\includegraphics[width=0.95\linewidth]{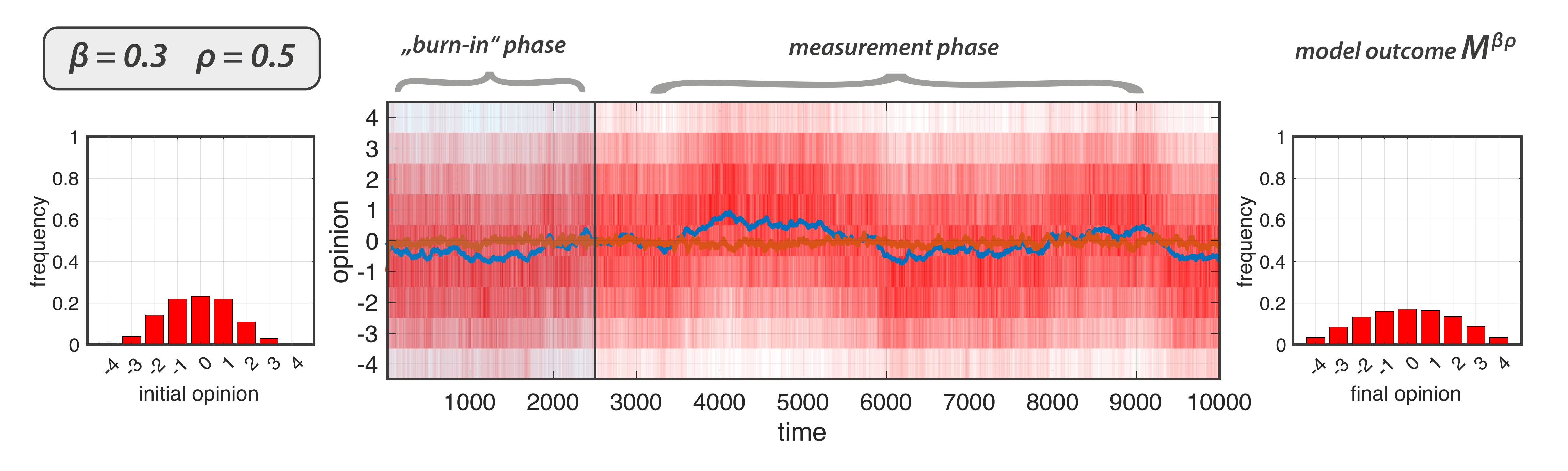}
	\includegraphics[width=0.95\linewidth]{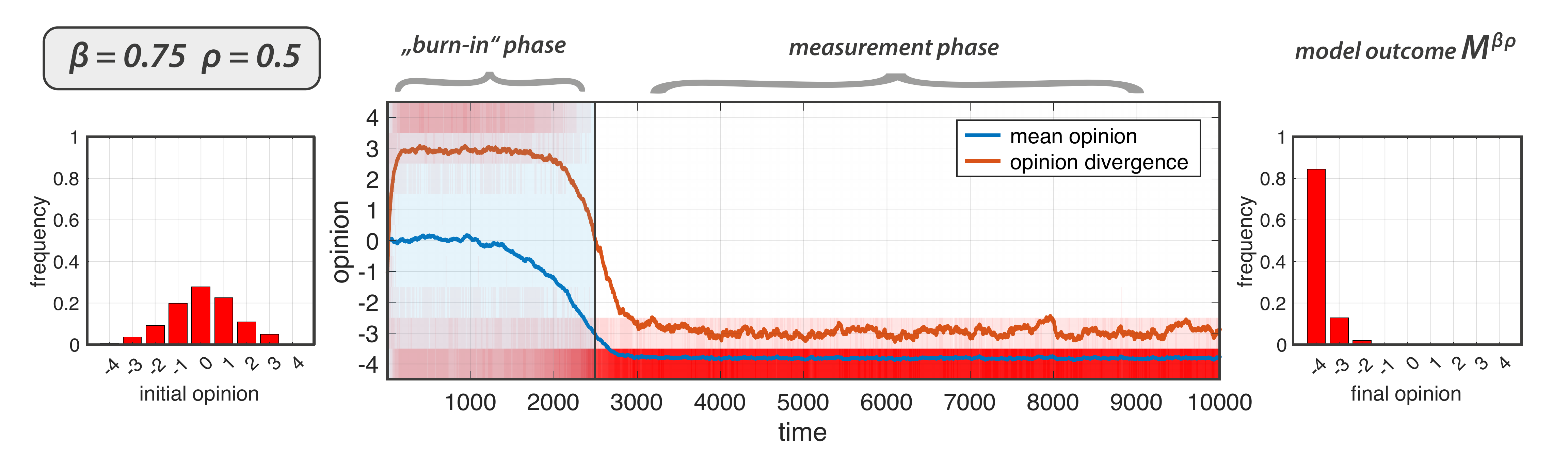}
	\includegraphics[width=0.95\linewidth]{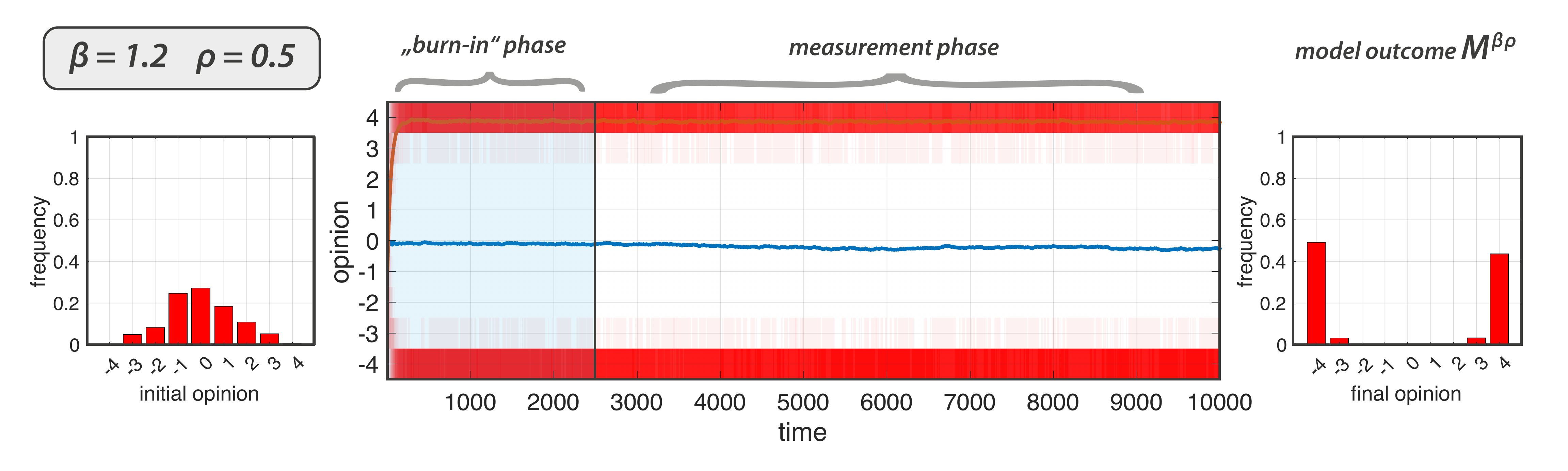}
	\caption{Three model run for $\rho = 0.5$ and $\beta = 0.3$, $\beta = 0.75$ and $\beta = 1.2$. 
	}
	\label{fig:run1}
\end{figure}

The model exhibits a rich dynamical behavior which we approach by three exemplary simulation runs. They are shown in Figure \ref{fig:run1}. 
We use $N = 500$ and $\rho = 1/2$ meaning that receivers get a random argument in 50 percent of the cases. The first 10000 time steps of the simulation are shown. From top to the bottom, we increase $\beta$ from $0.3$ to $1.2$. In all three case, the initial opinion distribution is shown on the left. In the center, the temporal evolution of the opinion distribution under the model is shown along with the respective mean opinion (blue) and the standard deviation (orange) as a measure of opinion divergence. Finally, at the right-hand side of the plots the opinion distribution averaged over the last 7500 steps is shown. We will treat this distribution as the model outcome (see below).

If $\beta$ is low (top) we observe that the opinion distribution remains centered at $o = 0$ and fluctuates slightly around it as time proceeds. In this case the dynamical behavior is driven by noise. As opposed to the model without bias and noise ($\beta = 0$ and $\rho = 0$) in which all agent approach the same opinion, the distribution remains rather broad and dynamic. For an increased $\beta = 0.75$ (middle) we observe a quick transition into an initial period of bi-polarization that persists for around 2000 steps. In the period around $2000 < t < 3000$ this state resolves in a collective choice shift towards a negative opinion ("extreme consensus"). Notice that with symmetric initial conditions the shift takes place to either side with equal probability. If $\beta$ grows large, this meta-stable bi-polarized regime persists throughout the entire 10000 time steps considered in Figure \ref{fig:run1}. This bi-polarized opinion regime may last very long, especially when $\beta$ increases further, but eventually also this realization will collapse into a one-sided consensual profile \citep[cf.][]{Banisch2021biased}. 

\subsection{What do we treat as a model outcome?}

The main aim of this paper is to provide a global picture on the opinion profiles that emerge from the model in order to identify parameters $\beta$ and $\rho$ for which empirical opinion data is reproduced. We treat the model as a data generating procedure and have to identify the typical opinion distribution generated by the model given the amount of biased processing $\beta$ and the relative importance of external unbiased news $\rho$.
In this, we have to be clear about what precisely we consider as the outcome of a model. Given the complex temporal patterns briefly discussed in the previous section, this is a non-trivial task that may involve decisions that are to some extent arbitrary. 

For the subsequent analyses in this paper we will closely follow previous work on systematic model analysis by \cite{Lorenz2021individual}. That is, we analyze the model with $N = 500$ agents and run simulations for 10000 steps. The first 2500 steps are neglected as a "burn-in phase" needed to reach a stationary profile. The remaining 7500 steps are considered as a measurement phase. 
The distribution of opinions over this period define the model outcome $\mathcal{M}^{\beta \rho}$ for a single simulation.\footnote{Notice that the Jensen-Shannon divergence used for empirical comparison involve the logarithm and it is usually favorable to have distributions with full support, meaning that there is a non-zero probability for any $o$.
We ensure full support by adding to the model output a single instance of a population with random opinions uniformly distributed in $\{-4,4\}$. The impact of this adjustment on the shape of the distribution is neglectable ($\approx 1/2500$).} The burn-in phase, the measurement phase and the resulting outcome distribution are shown in Figure \ref{fig:run1}. 

\subsection{Systematic simulations}

We perform a systematic computational analysis with respect to the strength of biased processing $\beta = 0,0.04,$ $0.08, \ldots , 1.4$ (36 sample points) and the relative influence of the external signal $\rho = 0,0.04,0.08, \ldots , 1$ (26 sample points).
For each of these $ 36 \times 26 = 936 $ sample points $(\beta,\rho)$, we run 25 simulations and store these 25 outcome distributions $\mathcal{M}^{\beta \rho}$ for subsequent analyses.
\revision{By considering multiple runs per  parameter constellation, we depart from the setting of \cite{Lorenz2021individual} who base their statistics on a single run.}

\subsection{Model and code availability}

Supplementary material for the reproduction of all analyses in this article on the Open Science Framework under \href{https://osf.io/5tz6g/}{osf.io/5tz6g/}. For interactive exploration (e.g. with the calibrated parameters), we provide an online version at \href{http://universecity.de/demos/ModelExplorer.html}{www.universecity.de/demos/ModelExplorer.html}, cited as \cite{BanischDemos} throughout the paper.


\section{Characterization and categorization of emergent opinion profiles}
\label{sec:analysis}

\subsection{Characterization of emergent opinion distributions}

\begin{figure}[ht]
\centering
\includegraphics[width=0.95\linewidth]{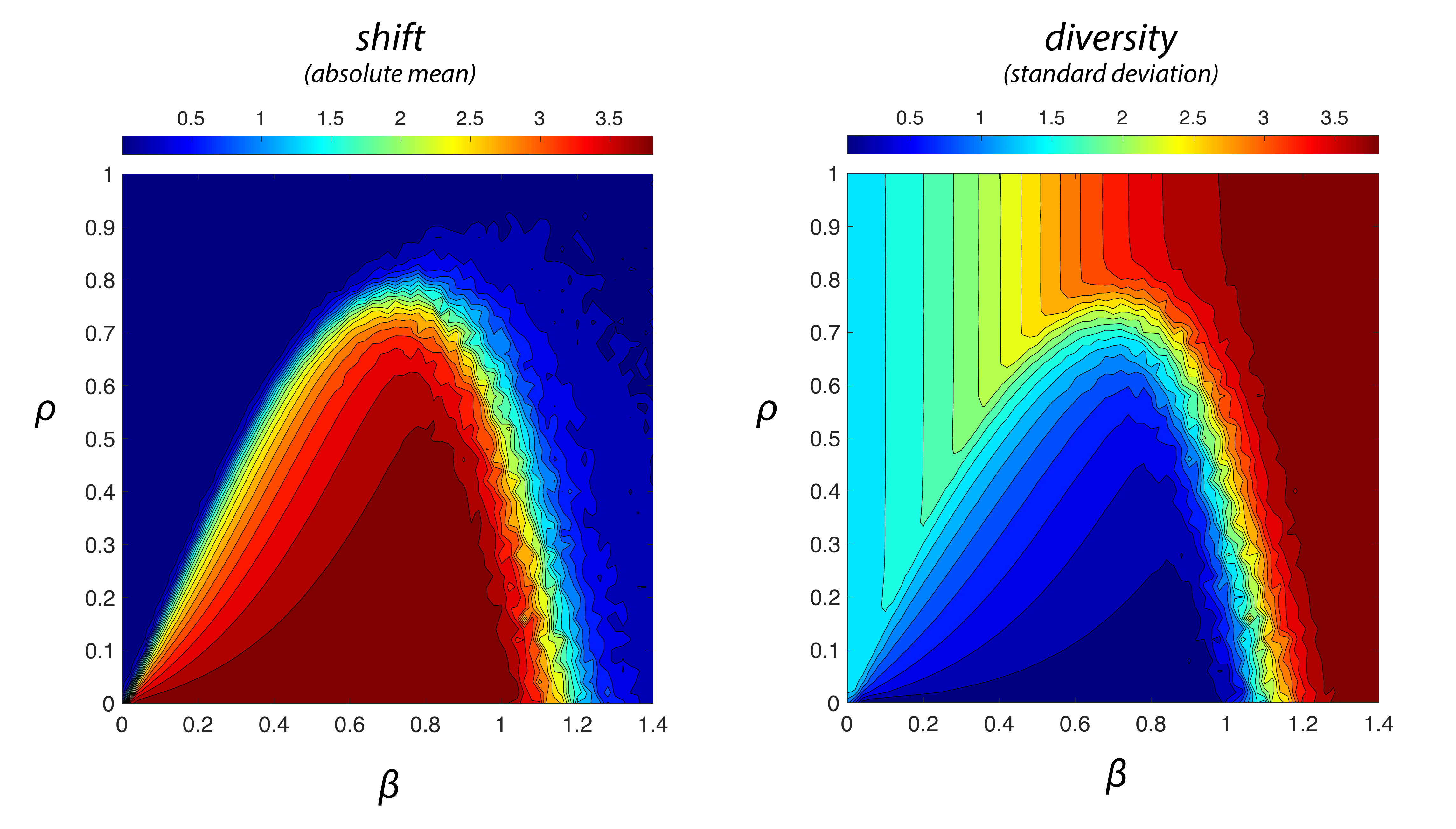}
\caption{The absolute mean value (shift) and the standard deviation (diversity) of the outcome distributions for $\beta \in [0,1.4]$ and $\rho \in [0,1]$. Red color indicates higher values, blue values close to zero. 
}
\label{fig:macromeanandstd}
\end{figure}

We can characterize the distributions $\mathcal{M}$ that emerge from the model by looking at the extent to which they are shifted toward one side and the amount of diversity or polarization. The latter is captured by the standard deviation of the distribution and allows to distinguish consensus profiles from diversified or polarized profiles. The sidedness of the distribution is captured by the absolute value of its mean which allows to distinguish between an extreme consensus on the one hand and bi-polarized and moderate consensus profiles on the other hand. In Figure \ref{fig:macromeanandstd}, the absolute mean value and standard deviation of the outcome distribution are shown for all parameter combinations $(\beta,\rho)$ in the considered ranges $(\beta \in [0,1.4],\rho \in [0,1])$.

\begin{figure}[ht]
\centering
\includegraphics[width=0.95\linewidth]{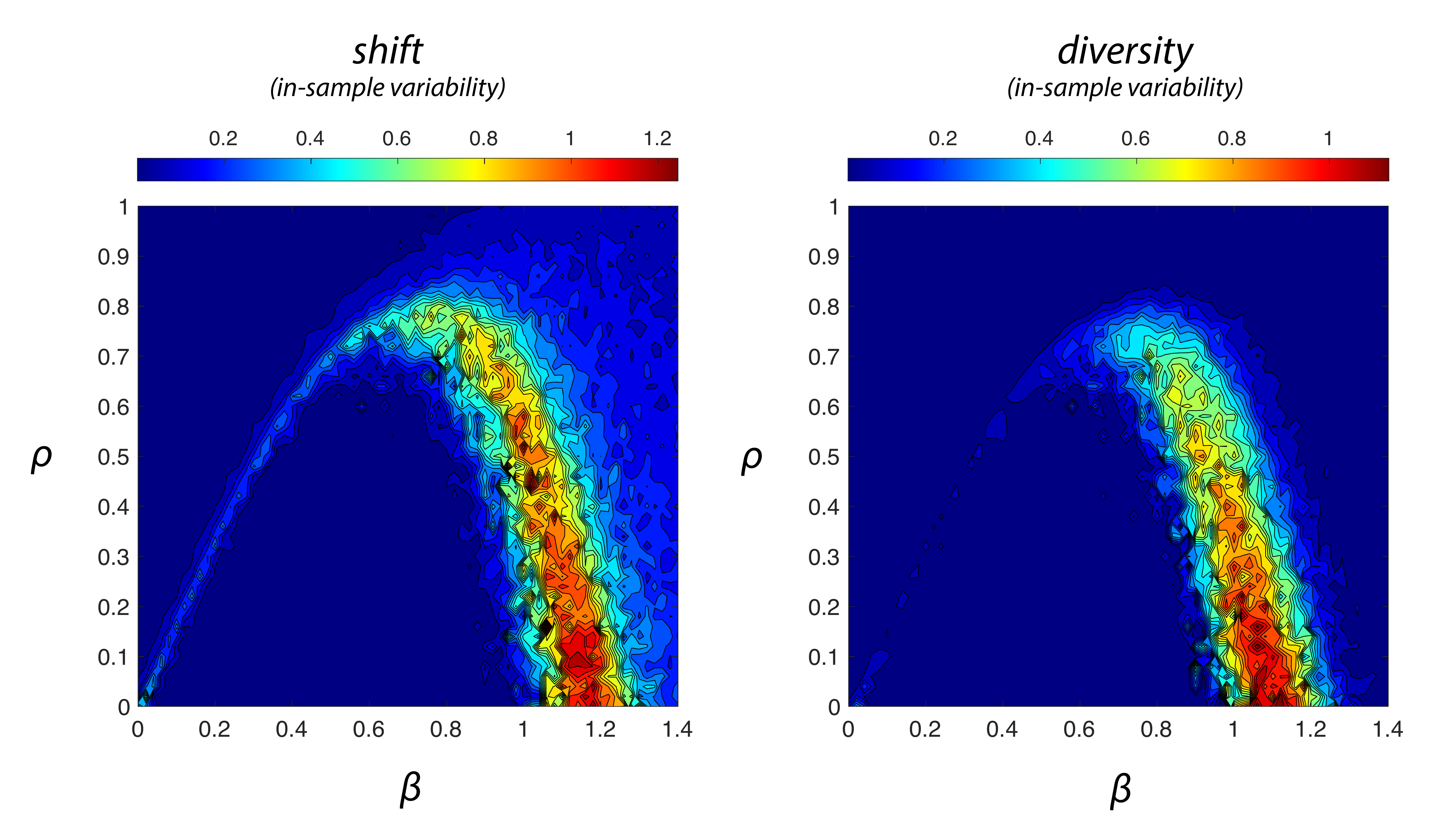}
\caption{Variability of the shift and diversity measure over 25 runs in each sample point $(\beta,\rho)$ for $\beta \in [0,1.4]$ and $\rho \in [0,1]$. Red color indicates higher values, blue values close to zero. 
}
\label{fig:macromeanandstdvar}
\end{figure}  
     
At the global scale, we observe a shift from a moderately diversified neutral distribution (close to normal) to a bi-polarized opinion distribution as $\beta$ increases. If the influence of the unbiased media channel is large ($\rho > 0.7$), there is a gradual increase of diversity while the shift remains close to zero. This indicates a soft transition from a normal, to a uniform to a more and more polarized distribution.
For values below ($\rho < 0.7$) we observe another intermediate opinion regime in which an one-sided consensus emerges. This regime is characterized by a large shift and low diversity.
As the impact of social argument exchange increases (diminishing $\rho$), this extreme consensus becomes a prevalent outcome over a wide range of biased processing strength $\beta$.
Note that in the limiting case $\rho = 0$ (only argument exchange) we recover the transition studied in \cite{Banisch2021biased}.

\subsection{Within-sample variability of shift and diversity}

Figure \ref{fig:macromeanandstd} shows the mean shift and diversity over 25 simulation runs with the same parameter combination $(\beta,\rho)$. We would typically expect that in parameter regions of transition from one qualitative model regime to another, there is higher variation in the model outcomes. To control for this effect, Figure \ref{fig:macromeanandstdvar} shows the in-sample variability measured as the standard deviation of both observables over the 25 runs.

The largest variability in both measures is observed in the transition region between one-sided consensus and bi-polarization.
This can be associated to the fact that bi-polarization is a meta-stable, transient phenomena in the model which becomes more persistent with increasing $\beta$ \citep{Banisch2021biased}.
In the band of high variability between $0.8 < \beta < 1.2$ and $\rho < 0.8$, a bi-polarized opinion profile may persist for many iterations, others may more quickly fall into the stable state of one-sided consensus. 
Then, while zero mean and high variance is assessed for the former, the latter is characterized by a large mean and low variance. 

Interestingly, the transition from a neutral opinion profile (normal or uniform distribution) to a strongly one-sided one is not associated with very large fluctuations. This indicates a rather sharp transition from moderate opinion profiles to one-sided, extreme profiles that reveal a clear collective preference for or against the issue.\footnote{This transition is an interesting object for future research.}

\subsection{Automatic categorization of different opinion regimes}

In order to classify the model outcomes into qualitatively different opinion regimes, \cite{Lorenz2021individual} propose an automated procedure based on these distributional measures including the mean, the variance but also more complex measures for the peakedness of a distribution.
In this paper we take a different approach and compare the outcome distributions to a series of stylized distributions using the Jensen-Shannon divergence (JS divergence).
The same approach has been used by \cite{Duggins2017psycologically} to show that the outcomes of his ABM come very close to opinion data on American politics.

Let's us denote a target opinion distribution by $\mathcal{D}$ and, as before, a model distribution with parameters $\beta$ and $\rho$ by $\mathcal{M^{\beta \rho}}$.
The JS divergence generalizes the Kullback-Leibler (KL) divergence and defines a distance between the two distributions by
\begin{equation}
    d_{JS}(\mathcal{M},\mathcal{D}) = \frac{1}{2} d_{KL}(\mathcal{M},M) + \frac{1}{2} d_{KL}(\mathcal{D},M) 
\end{equation}
with $M = \frac{1}{2}(\mathcal{M}+\mathcal{D})$. The KL divergence is defined as 
\begin{equation}
    d_{KL}(\mathcal{M},\mathcal{D}) = \sum_{o = -4}^{+4} \mathcal{M}(o) \log \frac{\mathcal{M}(o)}{\mathcal{D}(o)}
\end{equation}
where $\mathcal{M}(o)$ and $\mathcal{D}(o)$ are the model and target empirical frequencies of opinion $o \in \{-4,\ldots,4\}$ (9-point scale).
As in \cite{Duggins2017psycologically}, we will use the JS divergence to compare the model outcomes to empirical data in the next section.
In this section we propose to use it for an automated categorization of model outcomes into different qualitative regimes.

\begin{figure}[ht]
	\centering
	\includegraphics[width=0.6\linewidth]{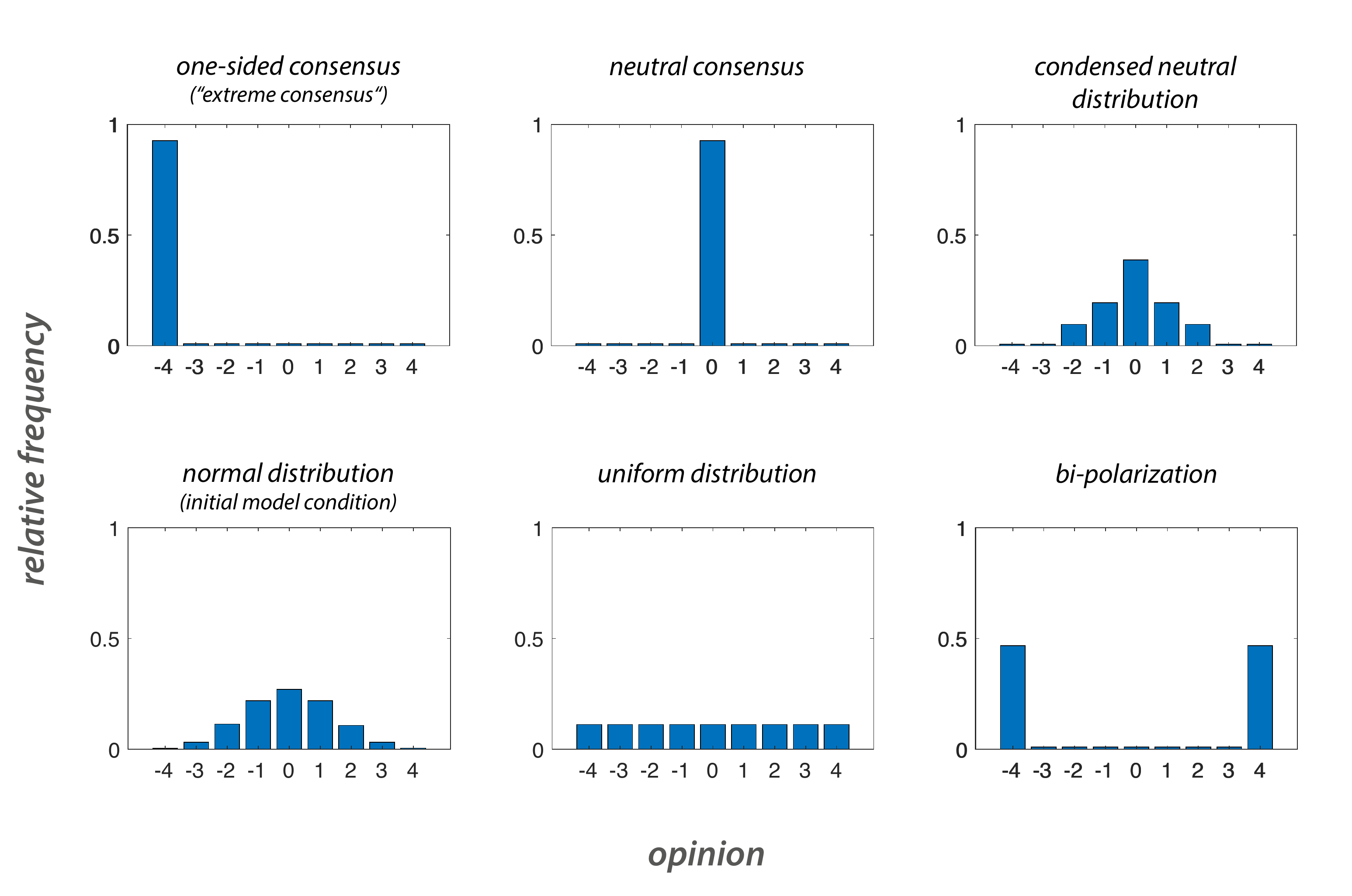}
	\includegraphics[width=0.38\linewidth]{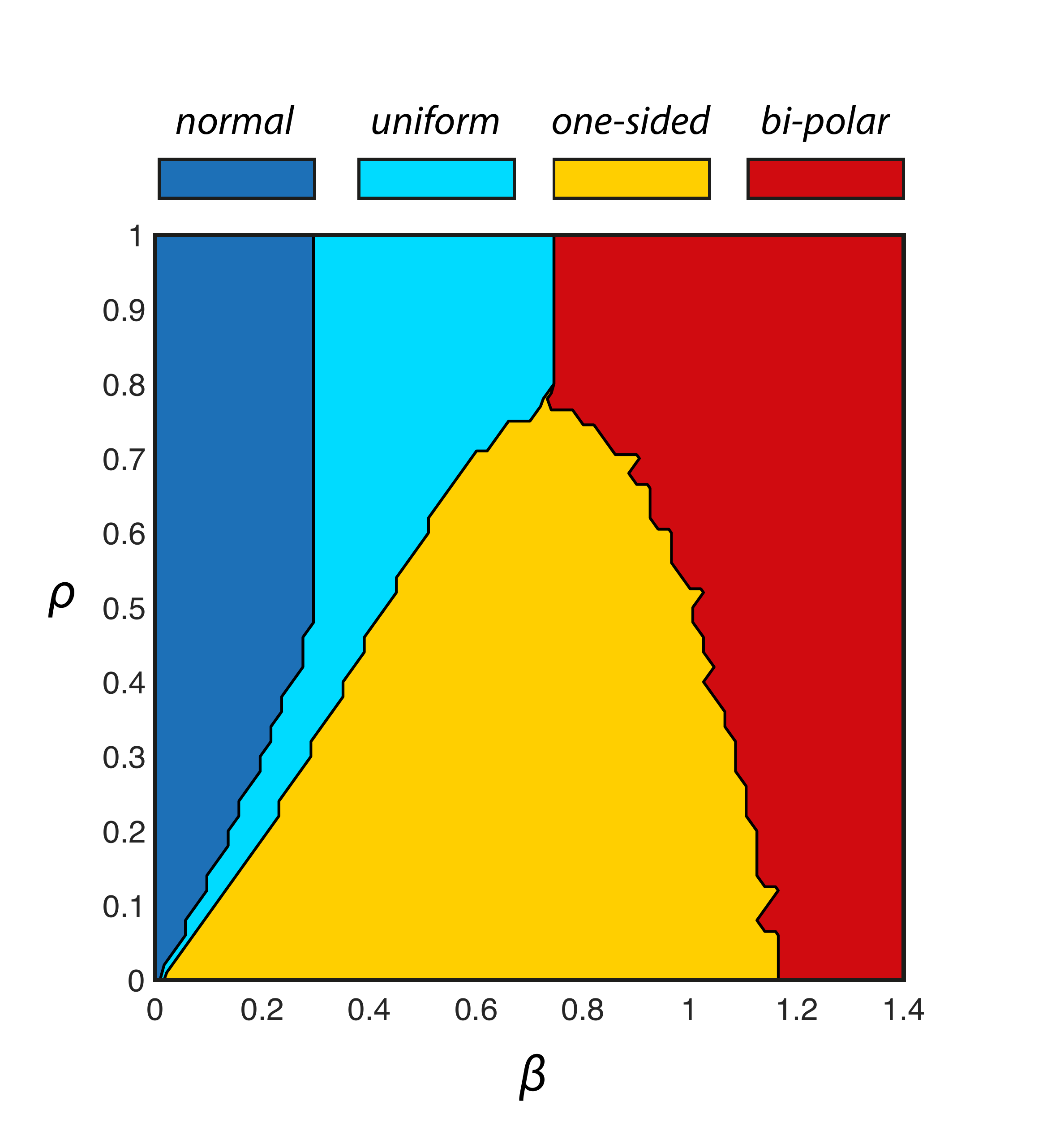}
	\caption{Comparison of the model outcomes with six idealized distributions shown on the left. 
	}
	\label{fig:macroregimes}
\end{figure}

For this purpose, we compute the JS divergence between a set of stylized distributions and the model outcomes $\mathcal{M}^{\beta \rho}$ for different $\beta$ and $\rho$.
The six idealized distributions ($\mathcal{D}^{i}$) used for comparison are shown on the left of Figure \ref{fig:macroregimes}.\footnote{Notice that a small base probability ($0.0093$) is assigned to all possible opinion values in $\{-4,4\}$. The peak in the consensus profiles, for instance, is $0.9259$, not one, and the remaining probability is equally distributed over the other opinion values. The technical purpose is to avoid taking a logarithm of zero in the computation of the JS divergence.}
For each parameter constellation $(\beta,\rho)$ we compute $d_{JS}(\mathcal{M}^{\beta \rho},\mathcal{D}^{i})$ between the model outcome and these six distributions.
We make use of the fact that the JS divergence is a proper distance and defines a metric in the space of opinion distributions (as opposed to the KL divergence on which it is based). This allows to identify which of these idealized distributions (indexed by $i$) is closest to the model outcome given some $\beta$ and $\rho$.
Notice that later on, we will use the same procedure to compare the model outcomes to the empirical survey data.

Over the entire parameter range $(\beta \in [0,1.4],\rho \in [0,1])$ only four of the six toy distributions are identified as "closest match" by this procedure. As already seen in Figure \ref{fig:macromeanandstd}, a large $\rho$ leads from an approximately normal distribution centered at the neutral opinion (dark blue) to one that is close to the uniform distribution (light blue) as $\beta$ increases. For $\beta > 0.7$ the distribution is already closer to the idealized bi-polarized distribution (red).
When $\rho$ is not too large $\rho < 0.7$, an entire range of model outcomes is clearly classified as one-sided consensus (yellow), compare Figure \ref{fig:macromeanandstd}. The shape of this yellow region reveals a surprising trend. When noise is introduced and its impact increases, opinion bi-polarization becomes more likely at lower values of $\beta$. This means that less biased processing can lead to polarization in the presence of unbiased external information. When biased processing is large, complete information diversity "enforced" by a constant inflow of an equal share of pro and con arguments does not prevent and may even foster bi-polarization.

Notice that the moderate consensus and the more condensed neutral distribution are nowhere closest to the model outcome.
The reason is that in the parameter space considered here moderate consensus is rare.\footnote{This is actually an intriguing observation given that until now the main question for opinion dynamics has been "how does polarization come about" \citep{Abelson1964mathematical,Axelrod1997dissemination,Flache2017models}. Similar findings concerning the marginality of moderate consensus have been made in \cite{Feliciani2021persuasion} and \cite{Lorenz2021individual} so that the focus in the future might shift back to "given that so many mechanism polarize, how can we remain moderate?"}
In fact, it only occurs at $\beta = 0$ and $\rho = 0$. As soon as $\beta$ becomes non-zero, opinions evolve into a one-sided consensus \citep{Banisch2021biased}.
When $\rho > 0$, opinions at a particular time may be condensed (only if $\rho$ is very small). 
But they drift through the opinion space as time proceeds, and taking the average distribution over a period of time will be close to the normal distribution. This points to a deficit of our definition of model outcomes.

Notice also that we do not differentiate whether one-sided consensus occurs at the negative or positive extreme. In the model both cases occur with equal probability. To deal with this fact, we compare the idealized distributions with the actual model outcome $\mathcal{M}^{\beta \rho}$ as well as with the inverse distribution mirrored at the neutral point at $o = 0$. The minimal JS divergence among these two cases is selected for the classification.

\revision{
\section{Validation approach}
\label{sec:validationapproach}

\subsection{Terminology}


The purpose of validation is to increase the trustworthiness of a theory or a model. Recently, it has been pointed out that validation is an ambiguous concept in the context of ABM and opinion dynamics in particular, because it may relate to very different activities \citep{ChattoeBrown2022today}. On the other hand, key terminology for establishing the credibility in simulation models more generally has been developed more than 40 years ago by \cite{Schlesinger1979terminology} and the "Society for Computer Simulation". These concepts, further developed by e.g. \cite{Sargent2010verification} and \cite{Sargent2017history}, are now established in the computer science literature on verification and validation (V \& V), and they may serve as an orientation for different validation activities in the context of ABMs \citep{David2009validation}. 

According to the V \& V literature, different activities to assess model credibility are related to different phases of the model development cycle: from the real phenomena of interest, to a conceptual model of the problem, to a simulation model, the outcomes of which are again confronted with empirical reality \citep[][]{Sargent2010verification}. This view entails the idea that simulation models are iteratively refined to capture intended phenomena with higher accuracy. On that basis, we can distinguish the following activities:
\begin{enumerate}
    \item 
    the objective of \emph{conceptual model qualification} is to show that the conceptual model is a qualified representation of the reality of interest. This process mainly concerns the assumptions at the conceptual level as it has to show >>that the theories and assumptions underlying the conceptual model are correct and that the model representation of the problem entity is "reasonable" for the intended purpose of the model<< \citep[][p.168]{Sargent2010verification}. 
    \item 
    \emph{computerized model verification} is the process which makes sure that a computer model accurately represents the developer's conceptual description. In the computational sciences this is also concerned with a mathematical study of the numeric algorithms involved into a simulation program, stability and sensitivity analyses and robustness tests. In the agent--based community, model to model comparison (M2M) \citep{Axtell1996aligning,Hales2003model} and model replication \citep{Edmonds2003replication,Rouchier2003re,Wilensky2007making} is a widely accepted method for model verification.
    \item 
    finally, \emph{model validation} is the >>substantiation that a computerized model within its domain of applicability possesses a satisfactory range of accuracy consistent with the intended application of the model<< \citep[][p.104]{Schlesinger1979terminology}. Model validation is done by carrying out different confirmation experiments to support the model by evidence, i.e., to confirm its assumptions on the model setup and its interaction rules. Therefore, the concept of \emph{empirical confirmation} plays an important role and will be the guiding principle in this work. 
\end{enumerate}

The established terminology of "verification" and "validation" is controversial in the simulation literature and beyond.
In their article on different philosophical positions on model validation, \cite{kleindorfer1998validation} argue that "the term validation, that is, 'to make valid', is already loaded with a philosophical commitment that a satisfactory model be rendered absolutely 'true'"  (p. 1089). This "either/or" logic is not only problematic from a philosophical viewpoint, it is also hardly feasible in practice. 
As pointed out prominently by Oreskes and colleagues in the context of Earth climate modeling, models cannot be validated in this absolute sense, since "in practice, few (if any) models are entirely confirmed by observational data, and few are entirely refuted" \citep[][p. 643]{Oreskes1994verification}.
However, with each confirming observation, the confidence in a model is increased, or, as \cite{Oreskes1994verification} put it: 
\begin{quote}
"[t]he greater the number and diversity of confirming observations, the more probable it is that the conceptualization embodied in the model is not flawed" (p.643).   
\end{quote}
Here we report two very different empirical confirmation tests, one at the micro and another one at the macro level, and argue that the consistency between the two with respect to biased processing is a viable sign for its empirical relevance in argument-based opinion models.

\subsection{Validation as consistent micro and macro calibration}

This paper deals with the empirical confirmation of argument communication \emph{theory}. Confidence in the theory is increasing if the outcomes of computational models devised on the basis of the theory satisfactorily fit with observational data in the intended application domain. As pointed out in \cite{Troitzsch2004validating}, >>[v]alidation of simulation models is thus the same (or at least analogous) to validation of theories<< (p.5). The paper encompasses two different empirical tests based one two different models both derived on the basis of ACT:
\begin{enumerate}
    \item 
    a microscopic model ($m^{\beta}$) of an artificial experiment to predict how individuals change their opinion when they receive arguments. This computational model of the experimental setting is conceptually aligned to the experimental part of the empirical study,
    \item 
    a macroscopic ABM ($M^{\beta,\rho}$) of collective opinion formation to study to what extent and under which parameters the emergent opinion distributions fit those observed in the survey part of the empirical study.
\end{enumerate}

With a micro model intended to capture individual attitude change by exposure to balanced arguments we show that a certain level of biased processing ($\beta$) explains short-term attitude changes with high accuracy. This has been the focus of the predecessor paper aiming at a refinement of ACT micro assumptions \cite{Banisch2021biased}. It confirms that a considerable level of biased processing is involved and renders previous neutral assumption \emph{invalid}. The macro model (ABM) intended to realistically capture empirical opinion distributions is the main focus of this paper. 

Following the suggestion to invent names for the involved validation methods \citep{ChattoeBrown2022today}, our approach could be named \textbf{MiMaCo} for "Micro and Macro Confirmation". However, the way in which the empirical confirmation tests are designed resembles more a parameter estimation approach. It is hence more closely related to empirical model calibration. Our claims about the empirical validity of ACT reside in the fact that the estimation results at the micro and the macro level are highly consistent. For this reason we could also refer to our approach as \textbf{CoMMCal} meaning "Consistent Micro and Macro Calibration". The overall validation pipeline is illustrated in Figure \ref{fig:validationscheme}.

\begin{figure}[ht]
	\centering
	\includegraphics[width=0.9\linewidth]{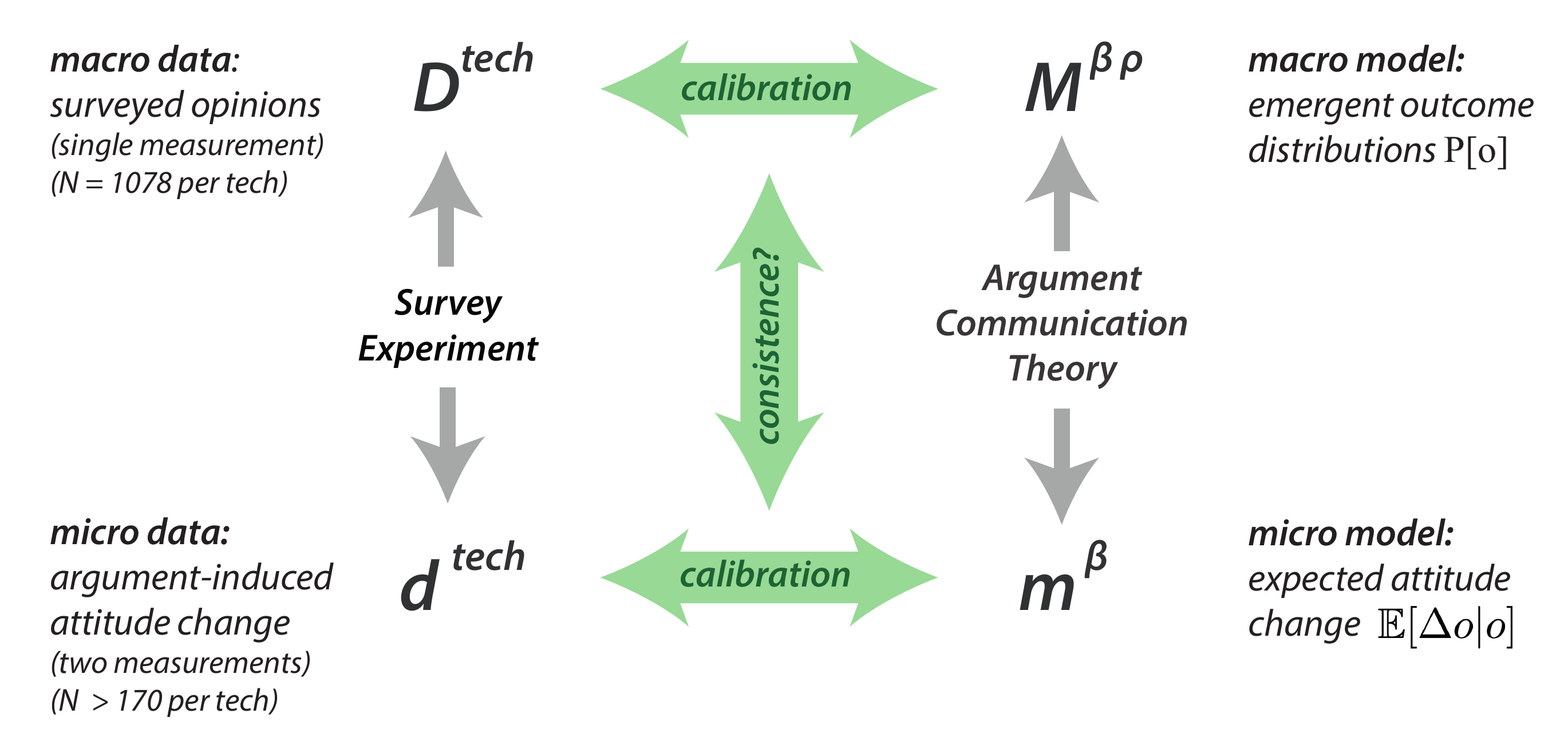}
	\caption{Overview of the validation approach adopted in this paper. The survey experiment (l.h.s.) provides two sources of data: (i.) micro data on individual opinion change induced by exposure to a balanced set of arguments for six experimental groups ($d^{tech}$), and (ii.) macro data on opinions on the six topics from the entire sample ($D^{tech}$). Based on ACT we devise a micro model with biased processing ($m^{\beta}$) and a macro model with biased processing and noise ($M^{\beta,\rho}$). Sampling through the parameter space, we identify the amount of biased processing $\beta$ and noise $\rho$ that matches the respective data, and confront the micro- and macro-level estimation results.}
	\label{fig:validationscheme}
\end{figure}

\subsection{The survey experiment}

In 2017, \cite{Shamon2019changing} performed a survey experiment to examine biased processing and argument-induced attitude change in the context of different electricity generating technologies.
The main aim has been to address the "puzzle of polarization" in social psychology. Namely, previous experimental work \citep[e.g.][]{Lord1979biased,Taber2006motivated,Taber2009motivated,Druckman2011framing,Corner2012uncertainty,Teel2006evidence} has lead to mixed evidence on whether exposure to balanced arguments leads to attitude moderation or polarization. Notice that in psychology attitude polarization refers to the individual-level tendency that subjects develop more extreme opinions after exposure to information. To address this puzzle in an experiment, an expert panel has developed a set of 84 arguments comprising 7 pro and 7 con arguments for six different electricity generating technologies (coal power stations, gas power stations, wind power stations (onshore), wind power stations (offshore), open-space photovoltaic, biomass power plants). 
Participants of the online survey were recruited from a voluntary-opt-in panel of a non-commercial German access panel operator.\footnote{The analytical sample consists of 1,078 persons who indicated to have a residential address (principal address) in Germany. Respondents’ average age in the analytical sample is 40.8 years (SD=15.7), and 49.3 percent of respondents are female, 49.4 percent are male, and 1.3 percent refused to classify their gender. Furthermore, 77.7 percent of the respondents had received a secondary school leaving certificate and 5.3 percent stated that they are employed in the energy sector.}
In the experiment, 1078 participants reported their opinion on all the six technologies ($D^{tech}$ in Figure \ref{fig:validationscheme}). Then they were randomly assigned to one of the technologies (N > 170) to receive the 14 balanced arguments tailored to that technology. Respondents were asked to rate, among other things, each argument’s persuasiveness as well as to state their perceived familiarity with each argument. The research design allowed to assess not only to what extent initial attitudes affect persuasiveness ratings of arguments but also to what extent respondents' initial attitudes change after the exposure to the balanced set of 14 arguments.


The structure of the experiment is highly compatible with ACT and provides micro and macro data needed to conduct the CoMMCal validation procedure. 
The setting provides subgroup data on individual attitude change ($d^{tech}$) and macro opinion data on all topics from the complete set of participants ($D^{tech}$). At the subgroup level it provides two subsequent opinion measurements for at least 170 individuals for six different opinion items (technologies). Opinions have been measured before and after exposure to the 7 pro and 7 con arguments and the time in between these two measurements was very short. The purpose has been to identify the effect of a single exposure treatment on opinion revision.

}


\revision{
\section{Empirical confirmation at the micro and the macro level}
\label{sec:multilevelconf}

\subsection{Micro-level calibration}

Out of the principles of ACT, \cite{Banisch2021biased} have developed a micro model directed at covering these short-term opinion changes. The main idea is to build an artificial version of the experiment which allows to predict how individual agents change their opinion when receiving an equal number of pro and con arguments. From the data, we obtain for each participant her previous opinion and the opinion change after exposure. In the micro model, agents initialized with a certain argument string (and hence opinion) are exposed to all arguments at once and we compute how many of the arguments are adopted according to the rules of ACT. Biased processing has been incorporated as a free parameter $\beta$ that governs how much this adoption probability is biased by the initial opinion (Eq. \ref{eq:biasedprocessing}). The expected attitude change for any given initial opinion can be analytically computed. 

This combination of experimental setup and artificial experiment allowed \cite{Banisch2021biased} to calibrate the modified version of ACT with respect to the biased processing parameter ($\beta$) by comparing participants’ observed attitude change in the empirical experiment with the model-based predictions on expected attitude change for their artificial twins, i.e., the agents. More precisely, \cite{Banisch2021biased} calculated the mean squared error (MSE) on the basis of the 1078 comparisons for different parameter values of $\beta \in [0, 1.2]$. The resulting function was u-shaped and exhibited a global minimum for MSE at $\beta \approx 0.5$ which lead the authors to the conclusion "that the argument adoption process refined with biased processing more appropriately captures argument-induced opinion changes" \citep[][p.9]{Banisch2021biased}. Furthermore, they replicated this procedure for each of the six different issues (i.e., different electricity generating technologies) separately and found that, along each of the six issues, a non-zero parameter value of $\beta$ improves the model predictions compared to models without processing bias.

These previous results are revisited in the final part of this section where we confront the micro- and macro-level estimation results.
}

\subsection{Macro-level calibration}

\subsubsection{Empirical opinion distributions}

\begin{figure}[ht]
	\centering
	\includegraphics[width=0.85\linewidth]{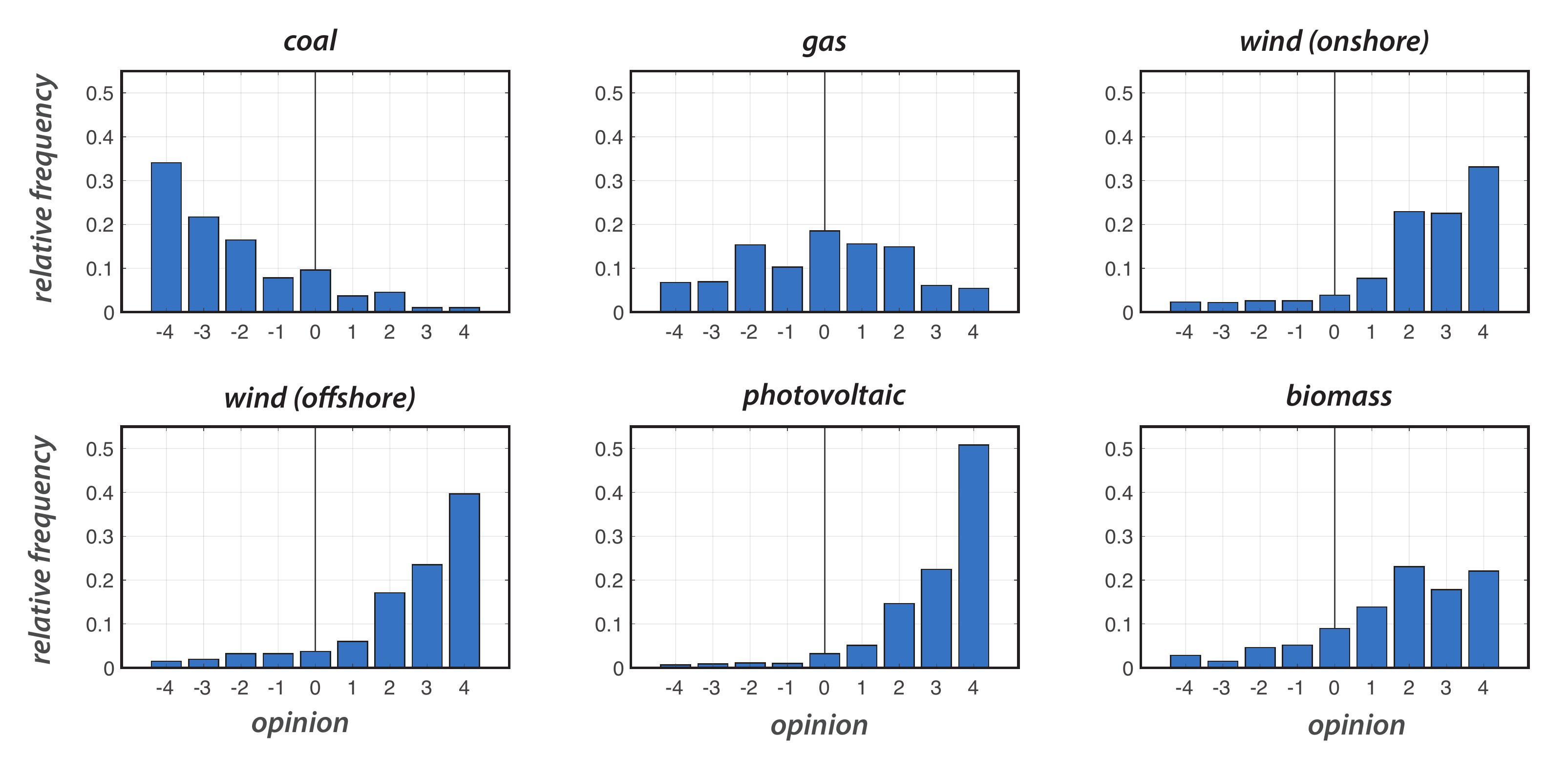}
	\caption{Surveyed opinions on six technologies (Setting 1: coal power stations; Setting 2: gas power stations; Setting 3: wind power stations (onshore); Setting 4: wind power stations (offshore); Setting 5: open-space photovoltaic; Setting 6: biomass power plants). 
	}
	\label{fig:empiricaldistributions}
\end{figure}

In the survey experiment of \cite{Shamon2019changing}, $N = 1078$ individuals reported their opinion on six different electricity-producing technologies. The six issues are coal power stations, gas power stations, wind power stations (onshore), wind power stations (offshore), open-space photovoltaic, and biomass power plants.
The survey operated with a 9-point attitude scale such that an opinion of -4 indicates a very negative opinion and +4 a very positive one.
In Figure \ref{fig:empiricaldistributions} the respective opinion distributions are shown.
These distributions will be the target of our analysis and we will assess how well the model fits this data under different parameter combinations of $\beta$ and $\rho$. 

In our sample\footnote{Participants of the online survey were recruited from a voluntary-opt-in panel of a non-commercial German access panel operator. The analytical sample consists of 1,078 persons who indicated to have a residential address (principal address) in Germany. Respondents’ average age in the analytical sample is 40.8 years (SD = 15.7), and 49.3 percent of respondents are female, 49.4 percent are male, and 1.3 percent refused to classify their gender. Furthermore, 77.7 percent of the respondents had received a secondary school leaving certificate and 5.3 percent stated that they are employed in the energy sector.}, there is a clear negative opinion tendency on coal power plants whereas wind power plants, open-space photovoltaic and biomass power plants are positively evaluated by the participants. The distribution on gas power is neutral but relatively broad. Given our previous analysis of the model behavior this already indicates that the model may fit these data (except gas) well with parameters that lead to a one-sided consensus (Figure \ref{fig:macroregimes}).
 
\subsubsection{Comparing model outcomes to survey data}

\begin{figure}[ht]
	\centering
	\includegraphics[width=0.95\linewidth]{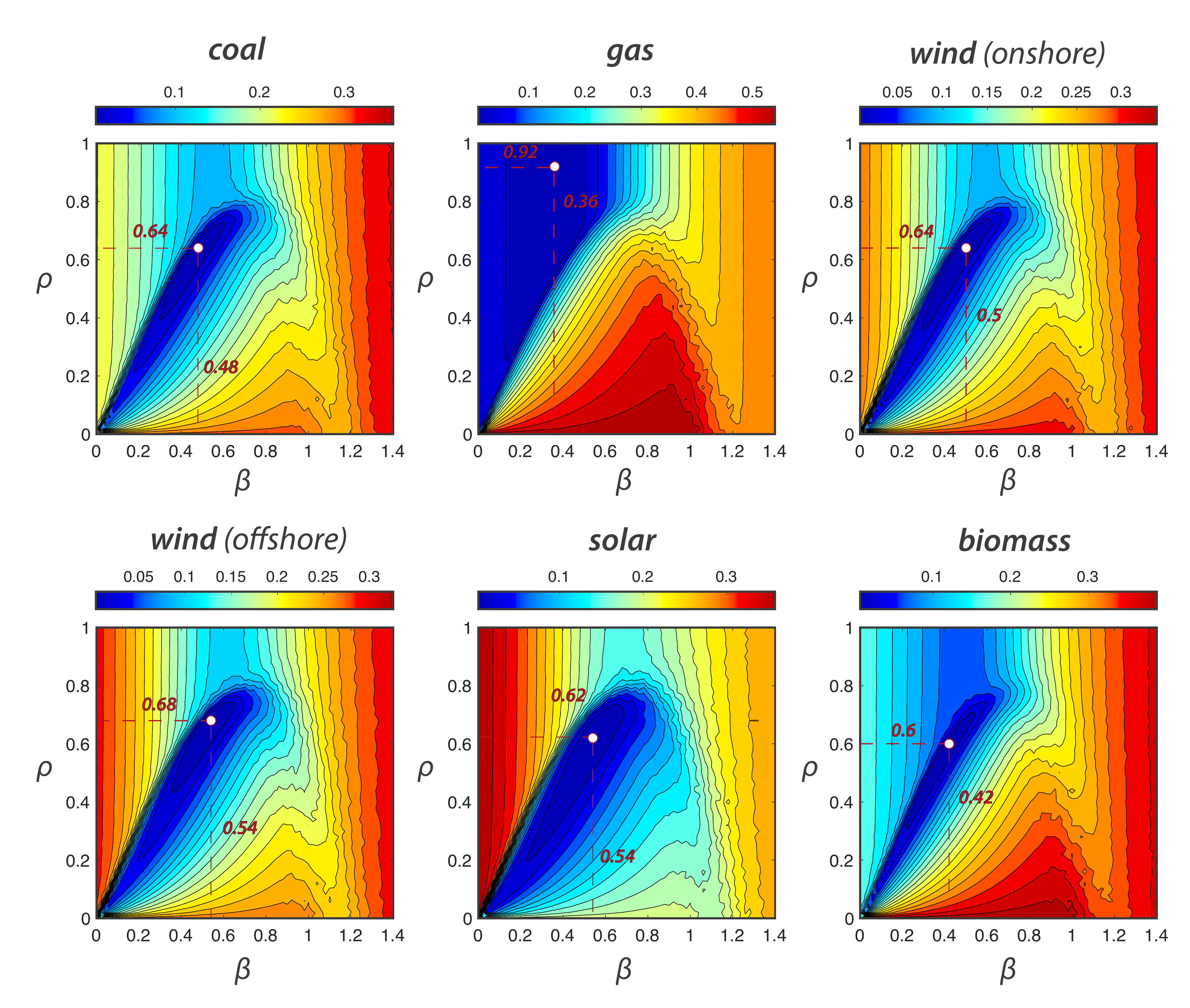}
	\caption{JS divergence between model outcomes and the six empirical distributions. Results are averaged over 25 within-samples per sample point $(\beta,\rho)$ with $\beta \in [0,1.4]$ and $\rho \in [0,1]$. The optimal parameter combination (minimal JS divergence) is shown by the white dot. The respective parameter values are superimposed to the plots.
	}
	\label{fig:fitJS}
\end{figure}

For each of the six technologies, we identify the regions in the parameter space $(\beta,\rho)$ of the model where the emergent opinion distributions match best with the empirical distribution of the survey.
For this purpose, we compute the JS divergence between the model outcome $\mathcal{M}^{\beta \rho}$ and the empirical target distribution $\mathcal{D}^{tech}$.
Notice that, as for the classification above, we are interested in the shape of the distribution and do not differentiate whether the opinion profile is drawn to one or the other side.
We therefore compute $d_{JS}$ also with respect to the mirrored model distribution and take the minimal divergence as the result. 
The test is performed on the computational data generated by the model sampling procedure described above (Section "Systematic simulations").
That is, we sample through the parameter space of the model varying the strength of biased processing $\beta = 0,0.04,$ $0.08, \ldots , 1.4$ (36 sample points) and the level of noice $\rho = 0,0.04,0.08, \ldots , 1$ (26 sample points).
We compute the JS divergence $d_{JS}(\mathcal{M}^{\beta \rho},\mathcal{D}^{tech})$ for all 25 within-sample outcomes independently and take the mean value as the final result. 
The mean JS divergence over different parameters is shown in Figure \ref{fig:fitJS}.

We hence adopt a statistical approach to model validation akin to calibration.
We treat the expected outcome of a model as a prediction $\mathcal{M}^{\beta \rho}$ of an empirical target distribution $\mathcal{D}$.
The model contains two parameters $\beta$ (strength of biased processing) and $\rho$ (noise level), and we "construct" a series of model predictions $\mathcal{M}^{\beta \rho}$ by sampling through the parameter space $(\beta,\rho)$.
Using the JS divergence, we asses the goodness of fit\footnote{We have also computed the log likelihood to compare model results with data. The results and inferred parameters are the same. In fact, the JS divergence is closely related to the log likelihood and serves the purposes of the analysis.} of $\mathcal{M}^{\beta \rho}$ and identify those parameter combinations $(\beta^*,\rho^*)$ that match best with the data. Formally,
\begin{equation}
    (\beta^*,\rho^*) = \arg \min_{(\beta, \rho)} d_{JS}(\mathcal{M}^{\beta \rho},\mathcal{D}^{}).
\end{equation}
In Figure \ref{fig:fitJS}, the respective "optimal parameters" are highlighted by a white dot and the numbers $(\beta^*,\rho^*)$ are shown in red.

The comparisons between model outcomes and the six empirical opinion distributions are shown in Figure \ref{fig:fitJS}. The analysis reveals that the model is capable of reproducing empirical distributions with high accuracy. In all cases there is a parameter combination $(\beta^*,\rho^*)$ with which the JS divergence is almost zero indicating an almost perfect fit. The analysis shows that these values are global minima within a relatively well-defined energy landscape (no local minima). This is very important from the point of view of model estimation because a unique, well-defined basin of minimal values is a signature that (i.) the model contains relevant information about the empirical case, and (ii.) that suitable model parameters are clearly discerned from suboptimal ones. 
For those technologies that are one-sided (all except gas power plants), there is a narrow band of parameter values with accurate model fit which is close to the transition region from a broad (uniform) distribution to one-sidedness (cf. Figure \ref{fig:macroregimes}).
These results are of course very similar as the empirical distributions are very close.
For gas, the best fit is obtained for high levels of noise in the opinion regime where the model continuously shifts from a normal-like to a uniform-like distribution.

\begin{figure}[H]
    \centering
    \includegraphics[width=0.95\linewidth]{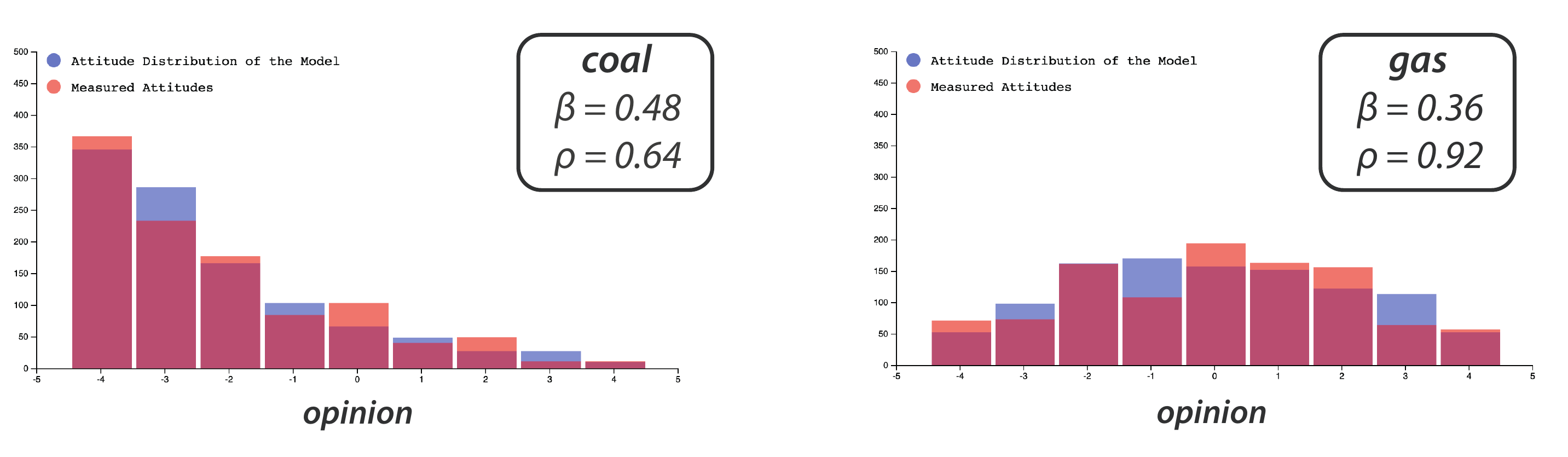}
    \caption{Empirical distributions for coal and gas along with a snapshot of the distributions of the model at the respective optimal $(\beta^*,\rho^*)$. The analysis is done with the \emph{online model explorer} \citep{BanischDemos} which uses $N = 1078$ agents. Distributions of opinions in a specific time step of the stationary phase are shown.
	}
    \label{fig:coalandgasfit}
\end{figure}

To further illustrate the capability of the model to generate the observed opinion distributions we have run experiments in our online application \citep{BanischDemos} using the respective parameters $(\beta^*,\rho^*)$ as an input.
For coal and gas, the empirical and model distributions are shown in \ref{fig:coalandgasfit}.
The reader is invited to test the goodness of fit for the other cases in the online tool.

\subsection{Consistency between micro- and macro-level estimation results}

\begin{figure}[ht]
    \centering
    \includegraphics[width=0.98\linewidth]{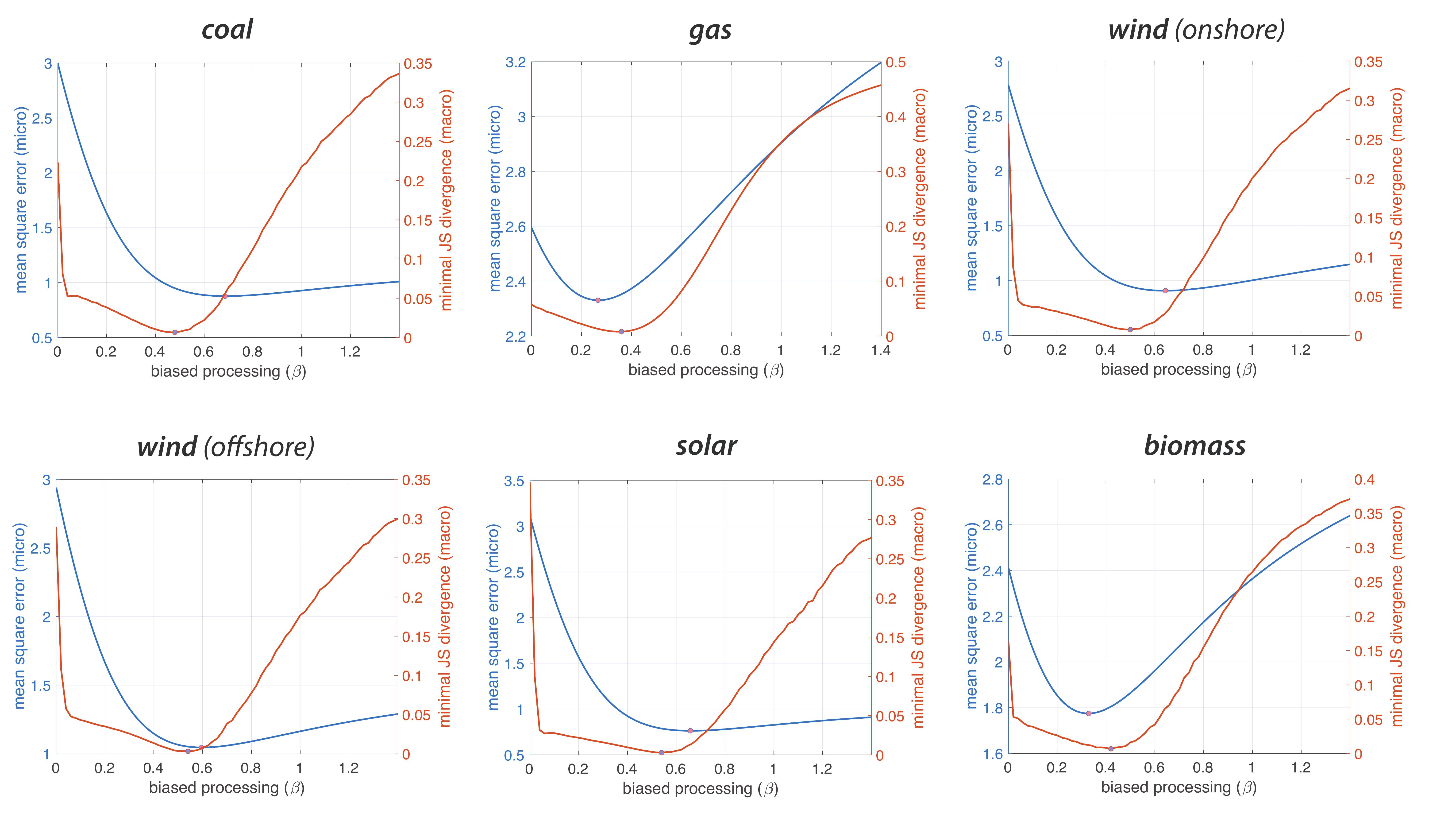}
    \caption{Comparison of estimation results for micro level and macro level data. The blue curves assess which amount of biased processing $\beta$ is most consistent with individual attitude change data \citep[adopted from][]{Banisch2021biased}. The orange curves show the minimal JS divergence between the ABM outcomes and the empirical opinion distributions taking the minimum values over $\rho$. The global minimum is highlighted in both cases, for the macro curves (orange) they correspond to the points highlighted in Figure \ref{fig:fitJS}.
	}
    \label{fig:mima_comparison}
\end{figure}

The previous section shows that the empirical distributions are matched with high accuracy for an ABM with moderate biased processors. This is consistent with the micro level estimation of $\beta$ in \cite{Banisch2021biased} where moderate values in between 0.25 and 0.7 have been identified.
In Figure \ref{fig:mima_comparison}, we show the estimation results for both analyses. 
The blue curves are taken from \cite{Banisch2021biased}. They show the mean squared error between experimentally observed opinion changes and the expected opinion change of artificial agents after reception of a balanced mix of arguments. The respective minimal value is highlighted.
The orange curves show the JS divergence based on Figure \ref{fig:fitJS}.
As the noise level $\rho$ has a non-trivial impact on the results, here we show the minimal JS divergence over all parameter values $\rho \in [0,1]$.
Noise mimicking an external information source is a property of the collective-level ABM has no direct correspondence at the level of individual agents. Taking the minimum of each column in the Figure \ref{fig:fitJS} ensures that the best possible match for a given $\beta$ is considered.

We observe in both cases moderate biased processing and the optimal values are not far apart. \revision{This means that a model calibrated at the micro level is capable of generating stationary opinion profiles very close to those observed in the survey part of the experiment.} To our point of view this justifies that a moderate level of biased processing $\beta$ enhances the empirical adequacy of argument-based models. 
The ABM provides a consistent link from micro data on individual attitude change to macro data in form of opinion distributions when agents are assumed to moderately favor consistent over inconsistent arguments. 

Notice that the two issues (gas and biomass) for which $\beta$ is significantly lower in the micro analysis are also lowest in the macro comparison presented in this paper. Beyond that, however, the ranking of optimal $\beta$'s does not match.
Hence, we do observe important differences.
Nevertheless, a central point can be made: moderate biased processing increases the empirical fit of the argument model both at the micro and the macro level.

\section{Concluding discussion}

Our paper uses a survey experiment to assess the validity of an argument-based model of opinion formation. Based on this experiment previous work has established biased processing (with strength $\beta$) as a viable micro assumption on how individuals change their opinion when exposed to new arguments \citep{Banisch2021biased}. In this work, we have concentrated on the macro-level validation by comparing model outcomes to the opinion profiles of the same surveyed population. For this purpose, the model has been extended by a noisy external signal the strength of which is governed by a second parameter $\rho$. The resulting ABM generates stationary opinion profiles that come remarkably close to the surveyed opinions in a well-defined region of the parameter space. Throughout the six empirical cases, moderate biased processing provides the best fit which is consistent with the micro analysis.

\revision{This consistency of micro- and macro level estimates in the CoMMCal setting entails the more typical calibration-validation approach in which a model calibrated at the micro level is shown to reproduce macro patterns. Consistent estimates of $\beta$ imply that the experimentally calibrated ABM is capable of matching the surveyed opinion distributions at the macro level for some noise level $\rho$. The approach adopted in this paper follows the route envisioned in the \cite{Flache2017models} review:
\begin{quote}
    "Calibrating models to resemble patterns observed in opinion surveys will be most fruitful if agent-based modelers at the same time assess to what extent those models that best fit macro-level patterns also contain assumptions that are compatible with empirical evidence available about micro-level processes of social influences and meso-structural conditions" (Par. 3.23)
\end{quote}
From that angle, the main methodological contribution of this research is to show that the mixture of survey experiments and agent-based modeling is well-suited to advance further towards empirically-grounded opinion models. Survey experiment provide data at both the micro- and the macro level, and in a CoMMCal consistent calibration setting the empirical value of an ABM can be judged with respect to how well it can explain both data.
}

\revision{ACT is particularly suited to enable such a comparison, because it provides a clear theoretical link between persuasion experiments, used to assess individual level change, and opinion models, used to assess emergent macroscopic effects from repeated interactions. Especially in opinion dynamics empirical measurement is a very hard task, and ACT relies on a formal experimental design rooted in argument persuasion research, which has -- in turn --} been the main inspiration for the original development of ACT \citep{Maes2013differentiation}. For this reason, a high conceptual alignment between empirical experiment and model has been obtained. In particular, the empirically measured and the modeled opinions both lie on a 9-point scale, allowing for a direct comparison without further transformation. The argument model with moderate biased processing relies on agents which adjust opinions realistically when put into experimental conditions, and is at the same time capable of reproducing the distributions of surveyed opinions. For the specific subject pool and the considered topics it provides a consistent explanatory link between micro-level data on individual attitude change and macro-level data on opinions. At both levels, it rules out the previous assumption of neutral argument processing \citep{Maes2013differentiation,Maes2013short,Maes2015will,Feliciani2021persuasion,Banisch2021argument,Keijzer2022complex}.

\revision{An intensive discussion of what validation may mean in the context of ABMs and opinion dynamics has emerged during the last two years \citep[see e.g.][]{ChattoeBrown2022today,Keijzer2022if,Neumann2023challenge}. While most researchers probably agree that validation involves quantitative comparisons of a model to experimental or survey data, also qualitative and interpretative approaches to model validation have been envisioned \cite[][]{Neumann2023challenge}, \citep[see also][]{kleindorfer1998validation}. We have argued (Section: \nameref{sec:validationapproach}) to view the question of validity not as} a binary one of either yes or no. Step by step we have to gather empirical support for a model, and to demarcate where it does not apply. This naturally means to become more specific in terms of the application context of the model and the phenomena it aims to explain. When it comes to model validation, there is no reason in seeking "grand theory of societal polarization", the empirical case studies pursued in validation work will enforce a smaller scope. However, once we have a model that can reasonably be confronted with data, each step -- whether successful or not -- can inform new experimental hypotheses and necessary model refinements to be tested and integrated in subsequent steps. In this way, we may widen the range of validity and approach what Robert K. Merton
called middle-range theories \citep{Merton1968social} of collective opinion dynamics. 


\revision{The above quote by \cite{Flache2017models} also points to current limitations our study that should be overcome in the future. We have not assessed structural conditions relating to social networks or patterns of media consumption. We have also not modeled meso-level structure.} The model studied in this paper is extremely simple with regard to the social composition of groups (no social network), the mechanisms of partner selection (no homophily), and the implementation of an external information channel mimicking an unbiased media source (just noise). In the light of the current experiment, we can justify these choices by the fact that we do not have sufficient knowledge on the relevant interactions and temporal processes within the surveyed population and their practices of media consumption. In that sense, we use randomness to account for factors we do not know. \revision{This strongly reduces the complexity of the social influence model to a single parameter $\rho$.} The aim is to provide a base line by showing that already this simple model may explain micro and macro opinion data if biased information processing is integrated into an argument-based ABM. 

\revision{Based on the two empirical confirmation tests that we report in this paper we are (i.) \emph{rather confident that moderate biased processing is a viable assumption} and should be included in future argument models. We are, however, (ii.) \emph{less confident in the influence processes modeled with the current ABM}, because other models might do equally well on fitting the data we have used. More and other data is needed to assess process validity, and to address the non-uniqueness problem related to "alternative model realisations" \citep[][p.24]{oreskes2001philosophical}.}

\revision{Most importantly, future iterations of the survey experiment should include opinion measurements some weeks after the experiment. At the moment, we have no data on opinion changes on the timescale of the ABM.
In order to address meso-level conditions, m}any aspects can be systematically varied in survey experiments, and there is an extensive body of previous experimental research on aspects such as source effects, expertise and argument quality as well as values and identities as mediators of opinion change \citep{Petty1986elaboration,Terry2001attitudes,Feldman2003values,Nelson2005values}.
\revision{To further increase confidence with respect to biased processing ($\beta$), the experiment should be reproduced on a different population and issues which show different degrees of polarization.
Argument communication theory makes precise predictions on the expected level of biased processing and the respective expected opinion change.}
At the macro level, one promising direction is to focus on group-level correlations between opinions on different issues. One the one hand, ideological patterns of opinion sorting are probably the most stable empirical signature of collective polarization \citep{Dimock2014political,dellaPosta2020pluralistic}. On the other hand, the emergence of opinion alignment has already been shown within the ACT modeling framework \citep{Banisch2021argument}, albeit without yet taking into account biased processing. 

\revision{With an ACT model that withstands these empirical tests,} we may more reliably address practical problems such as the impact of online social networks and algorithmic filters on polarization trends. 
In this regard, the model analysis has revealed a counter-intuitive effect of an ideal external channel that provides discussion groups a balanced mix of pro and con information -- quite the opposite of a filter bubble.
While intuitively one would expect that polarization reduces in such a scenario of perfect opinion diversity, more confrontation with balanced information may even be counter-productive and increase polarization tendencies \revision{\citep[cf. ][]{Deffuant2023regular}}. This effect will likely interact with homophily on which previous conceptions of personalized recommender systems have been based \citep{Maes2015will,Sirbu2019algorithmic,Keijzer2022complex}.
However, while the "complex link between filter bubbles and opinion polarization" \citep{Keijzer2022complex} may appear even more complex, we note that ACT with biased processing entails positive and negative influence at the phenomenological level of opinions \revision{\citep{banisch2023one}}. 
Aside from empirical model validation, future theoretical work on model synthesis is needed to clarify the relation between biased processing and these previous core modeling assumptions.

\endparano



% ...







\end{document}